\documentclass[sort&compress]{elsarticle}
\usepackage{graphicx}
\usepackage[colorlinks=true,urlcolor=blue,citecolor=blue]{hyperref}
\usepackage{booktabs} 
\usepackage[table]{xcolor}
\usepackage{import}
\usepackage{amsmath}
\usepackage{amssymb}
\usepackage{orcidlink}
\begin{document}
\begin{frontmatter}

\title{A Beamdump Facility at Jefferson Lab}

\hypersetup{pdfauthor={P. Achenbach, M.~Battaglieri, M. Bondi}}

\author[JLAB,JGU]{Patrick~Achenbach\,\orcidlink{0000-0002-6754-653X}\corref{cor}}%
\ead{patricka@jlab.org}

\author[GW]{Andrei~Afanasev}

\author[JLAB]{Pawel~Ambrozewicz} 

\author[TAU]{Adi~Ashkenazi} 

\author[CERN]{Dipanwita~Banerjee}

\author[INFN-Genova]{Marco~Battaglieri\,\orcidlink{0000-0001-5002-8771}\corref{cor}} 
\ead{battaglieri@ge.infn.it}

\author[JLAB]{Jay~Benesch} 

\author[INFN-Catania]{Mariangela~Bond\'i\,\orcidlink{0000-0001-8297-9184}\corref{cor}}
\ead{mariangela.bondi@ct.infn.it}

\author[ODU]{Paul~Brindza} 

\author[JLAB]{Alexandre~Camsonne}


\author[JLAB]{Eric~M.~Christy} 

\author[SBU,MIT]{Ethan~W.~Cline}
	
\author[JLAB]{Chris~Cuevas}

\author[JLAB]{Jens~Dilling}

\author[JGU]{Luca~Doria}
	
	
\author[York]{Stuart~Fegan}

\author[Messina]{Marco~Filippini}

\author[Messina]{Antonino~Fulci}

\author[INFN-LNF]{Simona~Giovannella}

\author[INFN-Genova]{Stefano Grazzi}

\author[DOE-ARPAE]{Heather~Jackson}

\author[JLAB]{Douglas~Higinbotham}

\author[JLAB]{Cynthia~Keppel} 

\author[JLAB]{Vladimir~Khachatryan}


\author[UHampton]{Michael~Kohl} 

\author[JLAB]{Hanjie~Liu}

\author[UMN]{Zhen~Liu}

\author[VT]{Camillo~Mariani} 

\author[CNU]{Ralph~Marinaro}

\author[URochester]{Kevin~McFarland} 

\author[FNAL]{Claudio~Montanari}

\author[FNAL]{Vishvas~Pandey}


\author[JLAB]{Eduard~Pozdeyev}

\author[JLAB]{Jianwei~Qiu} 

\author[JLAB,INFN-LNF]{Patrizia~Rossi}

\author[UPavia,INFN-Pavia]{Riccardo~Rossini}
	
\author[JLAB]{Todd~Satogata}

\author[JLAB]{Glenn~Schrader}

\author[PSI,UZH]{Adrian~Signer}

\author[OXY]{Daniel~Snowden-Ifft}

\author[INFN-Genova]{Marco~Spreafico}

\author[FNAL]{Diktys~Stratakis}

\author[UHampton]{Manjukrishna~Suresh}

\author[FIU]{Holly~Szumila}

\author[TAU]{J\'ulia~{Tena~Vidal}}

\author[LBNL]{Davide~Terzani}

\author[York]{Charlie~Velasquez}

\author[Canisius]{Michael~Wood}

\author[KEK,J-PARC]{Takayuki~Yamazaki} 

\author[JLAB]{Yuhong~Zhang}

\cortext[cor]{Editors.}


\address[JLAB]{Thomas Jefferson National Accelerator Facility, Newport News, Virginia 23606, USA}
\address[JGU]{Johannes Gutenberg University, 55128 Mainz, Germany}
\address[GW]{George Washington University, Washington, DC 20052, USA}
\address[TAU]{Tel Aviv University, Tel Aviv-Yafo 6997801, Israel}
\address[CERN]{CERN, 1217 Geneve, Switzerland}
\address[INFN-Genova]{Istituto Nazionale di Fisica Nucleare, Sezione di Genova, 16146 Genova, Italy}
\address[INFN-Catania]{Istituto Nazionale di Fisica Nucleare, Sezione di Catania, 95125 Catania, Italy}
\address[ODU]{Old Dominion University, Norfolk, Virginia 23529, USA}
\address[SBU]{Stony Brook University, Stony Brook, NY 11790, USA}
\address[MIT]{Massachusetts Institute of Technology, Cambridge, MA 02139, USA}
\address[York]{University of York, York YO10 5DD, United Kingdom}
\address[Messina]{Universit\`a degli Studi di Messina, 98122 Messina, Italy}
\address[INFN-LNF]{Laboratori Nazionali di Frascati dell’INFN, 00044 Frascati, Italy}
\address[DOE-ARPAE]{Advanced Research Projects Agency --- Energy, Department of Energy, USA}
\address[PSI]{Paul Scherrer Institute, 5232 Villigen, Switzerland}
\address[UHampton]{Hampton University, Hampton, Virginia 23668, USA}
\address[UMN]{University of Minnesota, Minneapolis, Minnesota 55455, USA}
\address[VT]{Virginia Tech, Blacksburg, Virginia 24061, USA}
\address[CNU]{Christopher Newport University, Newport News, Virginia 23606, USA}
\address[URochester]{University of Rochester, Rochester, New York 14627, USA}
\address[UPavia]{Universit\`a degli Studi di Pavia, 27100 Pavia, Italy}
\address[FNAL]{Fermi National Accelerator Laboratory, Batavia, Illinois 60510, USA}
\address[INFN-Pavia]{Istituto Nazionale di Fisica Nucleare, Sezione di Pavia, 27100 Pavia, Italy}
\address[UZH]{University of Z\"urich, 8057 Z\"urich, Switzerland}
\address[OXY]{Occidental College, Los Angeles, CA 90041, USA}
\address[FIU]{Florida International University, Miami, Florida 33199, USA}
\address[LBNL]{Lawrence Berkeley National Laboratory, Berkeley, CA 94720, USA}
\address[Canisius]{Canisius University, Buffalo, New York 14208, USA}
\address[KEK]{High Energy Accelerator Research Organization, Tsukuba, Ibaraki, 305-0801, Japan}
\address[J-PARC]{Japan Proton Accelerator Research Complex, Tokai-mura, Ibaraki 319-1195, Japan}

\begin{abstract} 
  This White Paper is exploring the potential of intense secondary muon, neutrino, and (hypothetical) light dark matter beams produced in interactions of high-intensity electron beams with beam dumps.   Light dark matter searches with the approved Beam Dump eXperiment (BDX) are driving the realization of a new underground vault at Jefferson Lab that could be extended to a {\em Beamdump Facility} with minimal additional installations. The paper summarizes contributions and discussions from the {\em International Workshop on Secondary Beams at Jefferson Lab (BDX \& Beyond)}. Several possible muon physics applications and neutrino detector technologies for Jefferson Lab are highlighted. The potential of a secondary neutron beam will be addressed in a future edition. 
\end{abstract}

\date{\today}

\end{frontmatter}

\setcounter{tocdepth}{3}
\tableofcontents

\section{Introduction}
%
Jefferson Lab is known for its comprehensive scientific program in hadron physics using the up-to-12~GeV electron beam from CEBAF, primarily involving fixed-target experiments. The interaction of the high-intensity electron beam with the dump generates intense secondary beams of muons, neutrinos, neutrons, and hypothetical dark matter particles~\cite{Battaglieri:2023gvd}. The research potential of a light dark matter (LDM) beam is being explored by the beamdump eXperiment (BDX) that has been approved by the Physics Advisory Committee of Jefferson Lab~\cite{BDX:2019afh}. The BDX-MINI experiment has already demonstrated the concept by successfully performing a LDM search with a dumped 2.1-GeV electron beam~\cite{Battaglieri:2022dcy}. The underground vault required for BDX could also host additional instrumentation for opportunistic experiments with secondary beams and could thereby expanded into a {\em Beamdump Facility} with additional installations. The {\em International Workshop on Secondary Beams at Jefferson Lab: BDX \& Beyond} (\href{https://indico.jlab.org/event/951/}{indico.jlab.org/event/951/}) explored the potential of intense muon and neutrino secondary beams at such a beamdump facility. A parallel study is underway to evaluate the potential of secondary neutron beams.

\section{The Beam Dump eXperiment (BDX) at Jefferson Lab}
%
\subsection{Probing Light Dark Matter at the Beamdump Facility}
%
Astrophysical observations suggest the possible existence of a new form of matter that does not interact directly with light: Dark Matter (DM). While the gravitational effects of DM are well established, its particle nature remains unknown despite extensive experimental efforts. Traditional searches have largely focused on the WIMP paradigm, which postulates weakly interacting massive particles (WIMPs) with masses above 1 GeV, but have so far yielded no conclusive evidence. This has motivated increasing interest in the exploration of DM in the MeV -- GeV mass range: light dark matter (LDM). Such a scenario is theoretically well motivated under the hypothesis of a thermal origin of DM: interactions with Standard Model (SM) particles would have enabled thermal equilibrium in the early universe, with the subsequent cosmic expansion freezing the DM abundance at its present value. For sub-GeV thermal DM, this mechanism requires the presence of new light mediators in the dark sector, capable of providing the necessary annihilation channels. Such mediators must couple to visible matter and be neutral under the SM $U(1)$ gauge group. Theoretical extensions of the SM that introduce new interactions and particles are highly constrained. Among the simplest extensions are scenarios involving new particles with a small electric charge (millicharged particles), light mediators such as dark photons, or more intricate hidden-sector structures. A commonly studied benchmark is the vector portal, where a dark photon ($A'$) kinetically mixes with the SM hypercharge~\cite{Holdom:1985ag}. In this framework, LDM can be produced through real or virtual decays of the $A'$.

On the experimental side, existing data, often collected in setups optimized for other physics goals, have been reanalyzed to set exclusion limits in the parameter space of the different models. A new generation of dedicated experiments with unprecedented sensitivity has already started data taking or is planned for the near future~\cite{Antel:2023hkf}. Present accelerator technology provides high intensity particle beams of moderate energy that are well suited for the discovery of LDM~\cite{Battaglieri:2017aum, Krnjaic:2022ozp}. In particular, electron beamdump experiments have been shown to have high sensitivity to LDM~\cite{Batell:2014mga, Andreev:2021fzd, Bondi:2017gul}. In these experiments, LDM particle pairs ($\chi {\bar\chi}$ pairs) are conjectured to be produced when the electron beam interacts with nucleons and electrons in the beamdump via an $A'$ radiative process ($A'$-\textit{strahlung}) or the annihilation (resonant and non-resonant) of positrons produced by the electromagnetic shower generated therein~\cite{Izaguirre:2013uxa, Marsicano:2018glj, Marsicano:2018krp}. A detector located downstream of the beamdump, shielded from SM background particles (other than neutrinos), could be sensitive to the interaction of $\chi$s.

\begin{figure}[t!]
    \centering
    \includegraphics[width=0.6\linewidth]{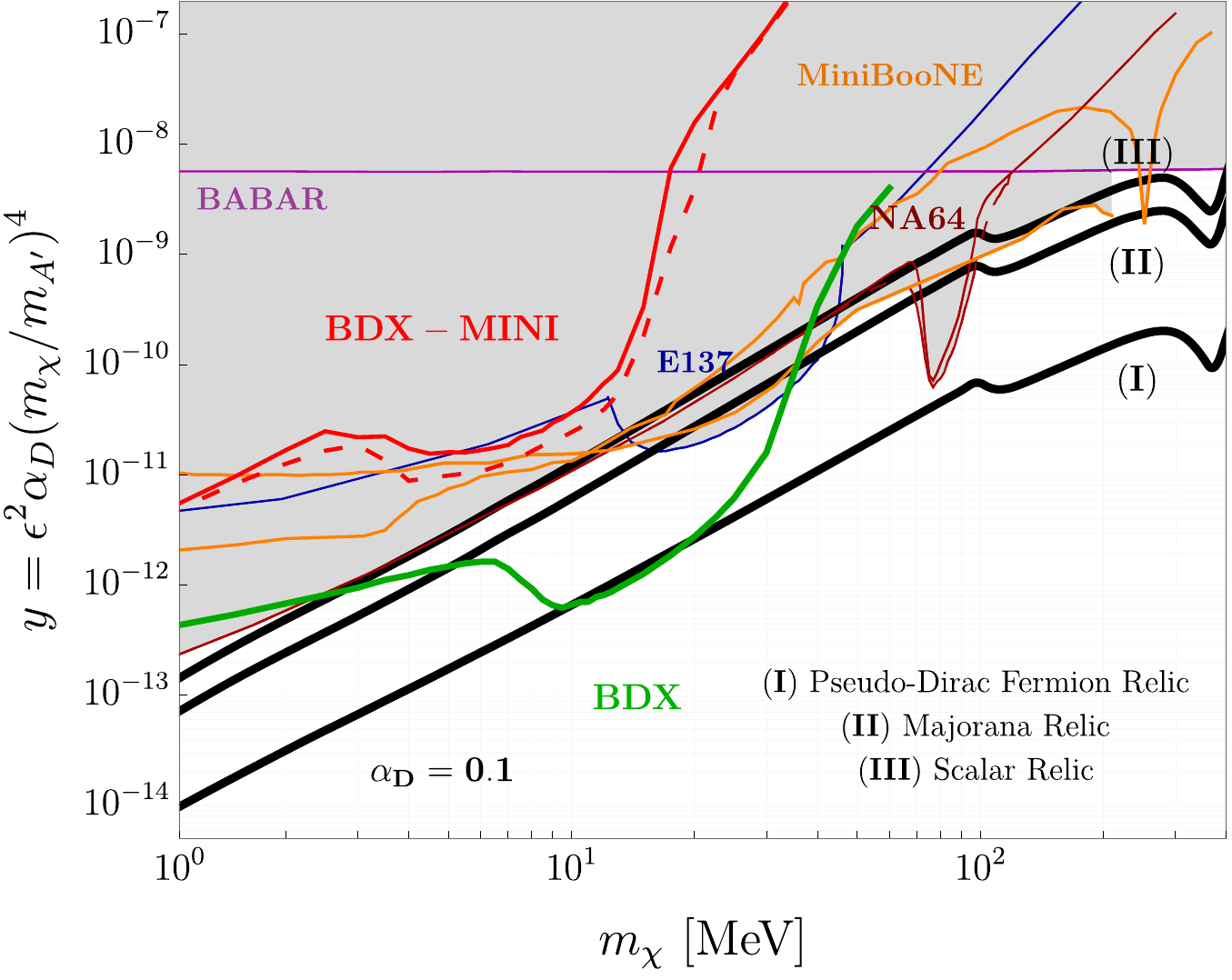}
    \caption{Projected BDX sensitivity for scalar LDM in the parameter $y$ versus mass $m_\chi$ plane, assuming $\alpha_D = 0.1$ and $m_{A'} = 3 m_\chi$ shown by the green exclusion limit.}
    \label{fig:reach_BDX}
\end{figure}

Jefferson Lab, providing one of the most intense electron beams worldwide, offers unique conditions for exploring the dark sector. BDX will be the first full-scale beamdump experiment explicitly designed for LDM searches. Planned to operate in parallel with the MOLLER experiment, BDX will exploit CEBAF’s high-current 11-GeV electron beam interaction with Hall A beamdump. A detector will be placed about 20\,m downstream in an underground vault. The BDX detector consists of an electromagnetic calorimeter, made of approximately one-cubic-meter of BGO and PbWO$_4$ crystals, surrounded by plastic scintillators and lead layers serving as active and passive veto systems. The characteristic signature of $\chi$--electron scattering will be an electromagnetic shower with deposited energy above 10 MeV, efficiently detectable by the calorimeter. Given the small production cross sections and interaction rates, a large number of electron-on-target (EOT) is required. With an expected accumulated charge of $\sim$10$^{22}$ EOT collected during the experiment lifetime, BDX will cover previously unexplored regions of the parameter space. BDX will test the model of LDM production via the $A'$ decay with a sensitivity exceeding up to two orders of magnitude the reach of existing experiments, see Fig.~\ref{fig:reach_BDX}. Other scenarios, including LDM interaction mediated by leptophilic mediators~\cite{Marsicano:2018glj}, ALPs~\cite{BDX:ALPs},  and inelastic LDM will be tested as well using the same detector.

\subsection{Probing Millicharged Particles at the Beamdump Facility}
%
Millicharged particles can also be copiously produced in high-intensity electron beamdump experiments. The BDX detector is optimized for electron-recoil searches and operates at a relatively high energy threshold $\sim\mathcal{O}(100~\text{MeV})$. As a consequence, its sensitivity to millicharged particles with $m < \mathcal{O}(10~\text{MeV})$ is very limited. To overcome this, and following the strategy outlined in~\cite{Essig:2024dpa}, BDX can be complemented with ultralow-threshold Skipper-Charge Coupled Devices (Skipper-CCDs)\cite{Tiffenberg:2017aac}.
A large flux of millicharged particles can be produced in the dump, allowing even a modest $2\times 14$ array of Skipper-CCDs to exceed the sensitivity of all existing searches for millicharged particle masses below 1.5~GeV, and to be either competitive or world-leading when compared to other proposed experiments. 
The potential of Skipper-CCD technology for accelerator-based dark-sector searches has already been demonstrated by the SENSEI collaboration, which set limits using a single device opportunistically located downstream of the NuMI beam at Fermilab~\cite{SENSEI:2023gie}, and is being further pursued by Oscura with a staged program towards kg-scale detectors~\cite{Oscura:2023qch}. Additional searches have also been carried out in reactor environments~\cite{CONNIE:2024off}.

\begin{figure}[!t]
  \centering
  \includegraphics[width=0.495\linewidth]{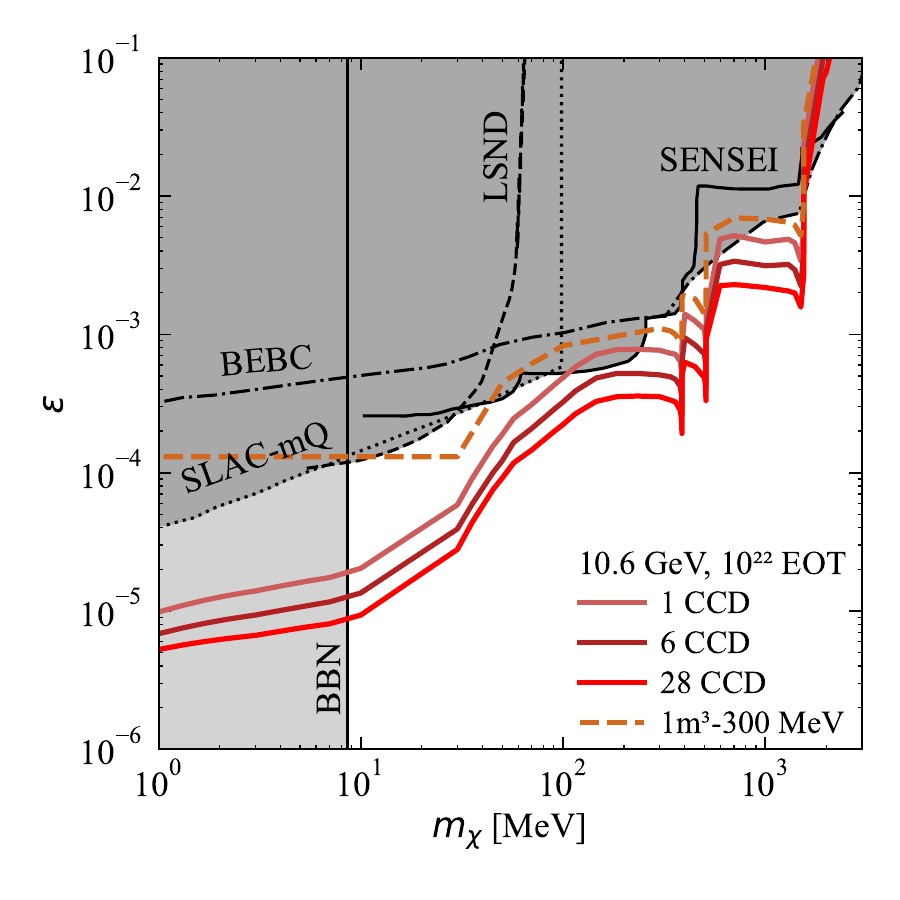}
  \includegraphics[width=0.495\linewidth]{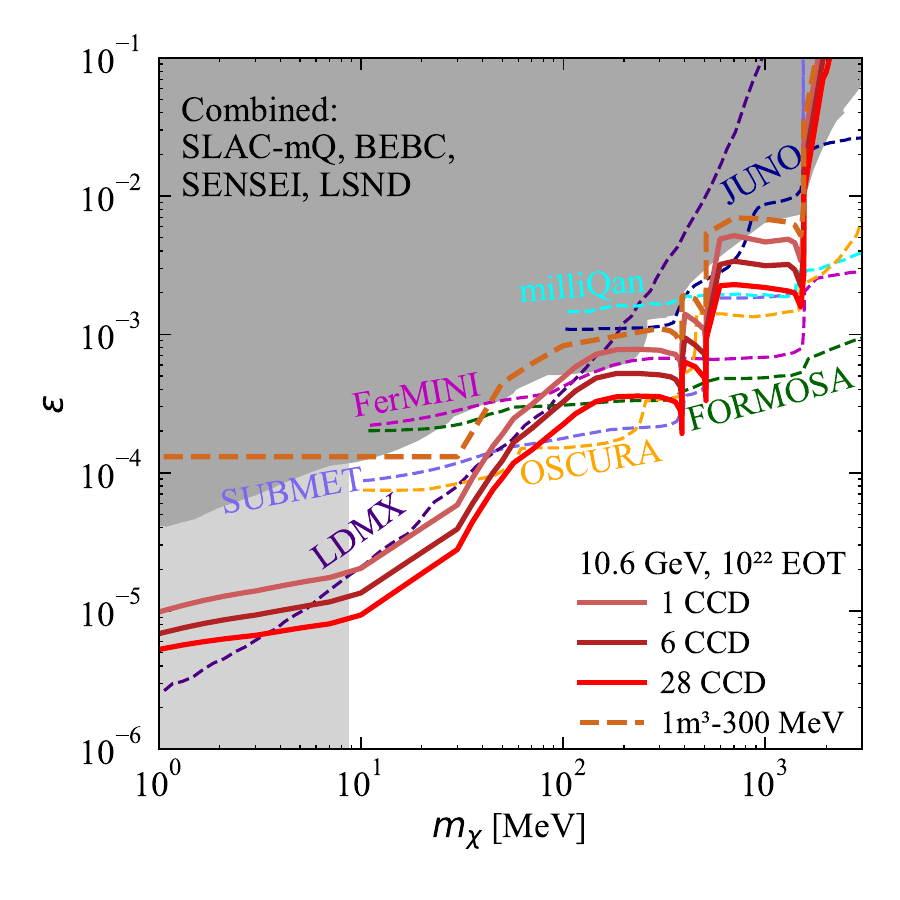}
  \caption{{\bf Left:} Projected BDX sensitivity to millicharged particles in the millicharge $\varepsilon$ versus mass $m_\chi$ plane of a Skipper-CCD detector array placed behind the proposed BDX detector. {\bf Right:} Same as left plot, but other proposed experiments are included as dashed curves.}
  \label{fig:pure_mCP_plot}
\end{figure}

Integrating Skipper-CCDs into BDX would provide a powerful and versatile complement to the calorimetric detector, extending the physics reach to a broad class of light and feebly interacting particles, including millicharged particles, as shown in Fig.~\ref{fig:pure_mCP_plot}, and dark photons. This idea has been formalized in a Letter of Intent submitted to Jefferson Lab's Program Advisory Committee (PAC53), which recognized its potential and recommended that the proponents submit a full proposal.

\section {Baseline Design for the Beamdump Facility}
\begin{figure}[htbp]
  \centering
  \includegraphics[width=0.96\textwidth]{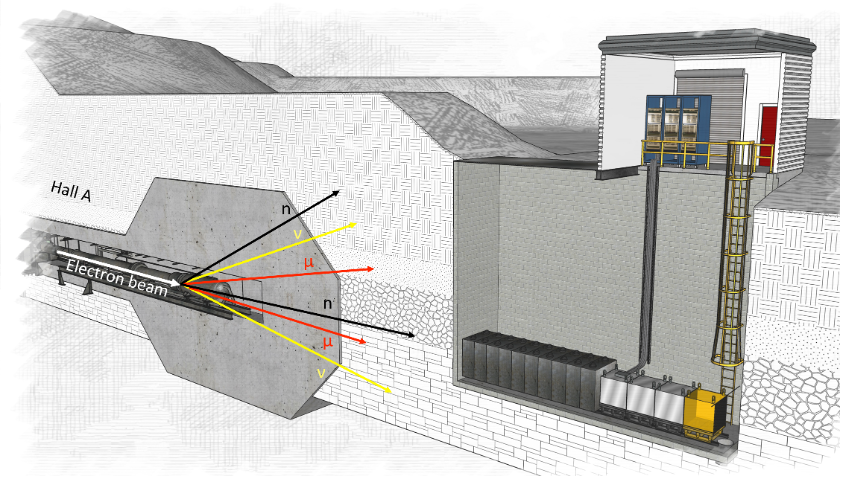}
  \caption{Sketch of the electron beamdump in Hall A (left) where intense beams of secondary neutrons, neutrinos and muons are produced. The beamdump facility (right) would provide many science applications for these beams. }
  \label{fig:bdx_facility}
\end{figure}

%
The BDX detector will be installed in a dedicated facility located downstream of the Hall-A beamdump. An artist's conception is shown in Fig.~\ref{fig:bdx_facility}. The main features of the beamdump facility include a vault divided into two areas: the upstream section, which serves as shielding and has no ceiling, and the downstream section, which houses the BDX detector and is covered by an above-ground building.

The vault is designed as a reinforced underground structure. The floor elevation in Hall~A is 3\,ft\,6'' above sea level, while the beamdump floor is set at 10\,ft. The vault will have a floor elevation of approximately 10\,ft\,6''. The interior of the vault will be positioned 42\,ft\,4'' from the interior face of the beamdump (24\,ft\,4'' from the exterior of the shielding). Since the vault floor is significantly below the water table, the design includes waterproofing and a sump pump system to remove accumulated water. The weight of the concrete structure will also be sufficient to counteract buoyant forces and prevent flotation due to groundwater pressure.

\begin{figure}[htbp]
  \centering
  \includegraphics[width=\textwidth]{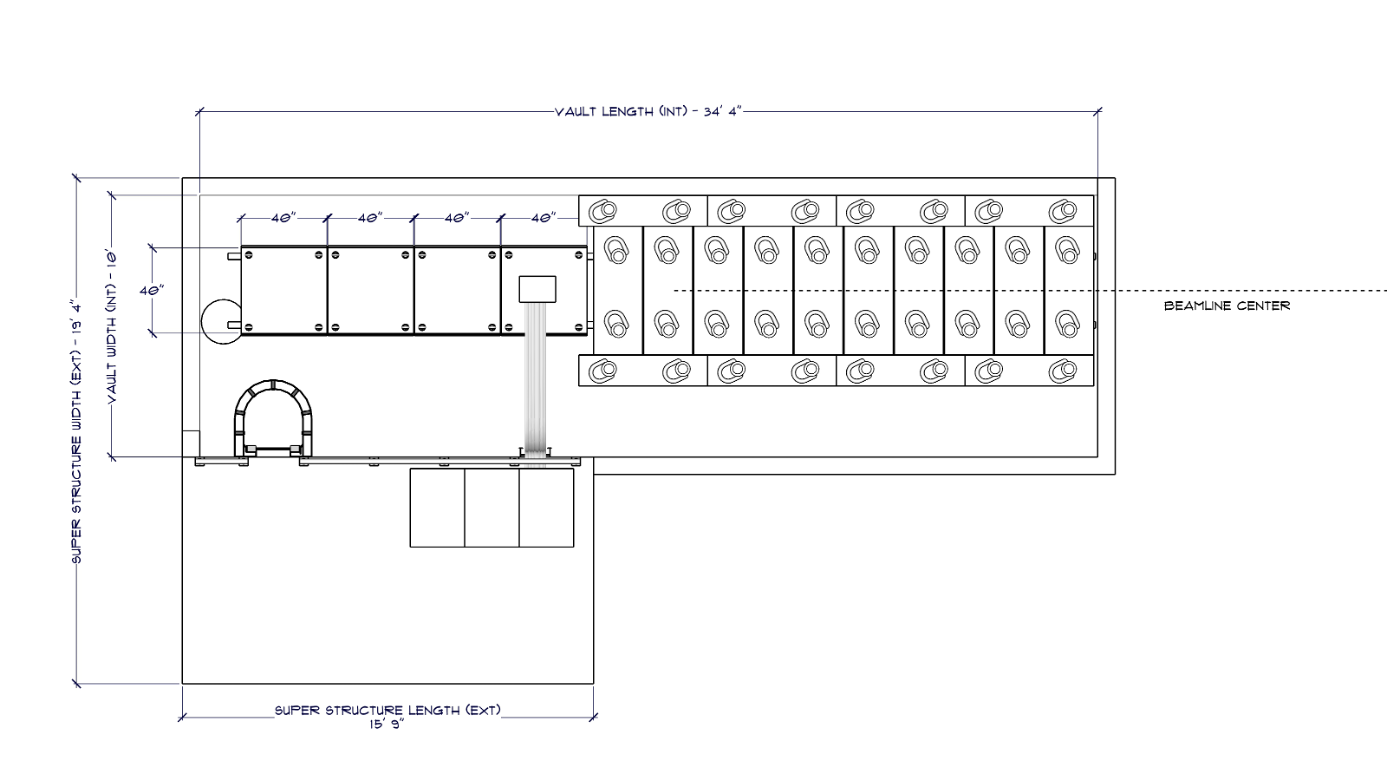}
  \caption{Dimensions of the underground vault of the beamdump facility. The drawing indicates the shielding and calorimeter blocks for BDX.}
  \label{fig:bdx_vault}
\end{figure}

The interior dimensions of the vault are 34\,ft\,4'' in length and 10\,ft in width, providing sufficient space for shielding and detector installation while allowing safe personnel access and navigation. The plan is shown in Fig.~\ref{fig:bdx_vault}. The back section of the vault, measuring 15\,ft\,9'', lies beneath the superstructure, which houses electronics and provides ladder access. The vault depth is approximately 28\,ft, although the final depth will be determined to ensure precise alignment with the Hall~A beamline.

The vault floor will be constructed as a monolithic slab of uniform thickness. It must accommodate 1,070\,ft$^3$ of shielding steel weighing approximately 524,000\,lb (262\,tons). This requires a load-bearing capacity of 2,550\,lb/ft$^2$. Due to the vault’s long walls, additional reinforcement may be necessary to prevent buckling; this can be achieved using a partial interior partition or exterior supports such as soldier piles. An access panel will be installed adjacent to the superstructure, minimizing or eliminating the need for cranes inside the structure. Shielding blocks and detector components will be lowered through the access panel onto supporting rails mounted on the vault floor. Once positioned, mechanical jacks or winches will allow sliding the components into their final locations. Personnel access will be provided by a ladder equipped with a protective cage. Due to the ladder’s height, additional fall-protection systems may be required. The superstructure, extending above the rear portion of the vault, has exterior dimensions of 19\,ft\,4'' in length and 15\,ft\,9'' in width. The interior height is 12\,ft, although usable clearance may be reduced locally by structural girders.  

Because of the vault overlap, usable floor space within the superstructure is 8\,ft\,8'' long and 15\,ft\,9'' wide. A personnel access door and a roll-up door (10\,ft tall, 8\,ft wide) will be provided. The superstructure will also host the electronic readout systems, with electrical power and minimal cooling supplied as required. The superstructure will receive 480\,V, three-phase electrical service, with step-down transformers for 110\,V loads. The main feeder will include surge protection, and a robust grounding system will be implemented using busbars in both the superstructure and vault. The electrical system will also support lighting, cooling, and convenience outlets. Air conditioning will regulate the superstructure environment at 75$^\circ$F. The vault will not be cooled, though an isolating barrier may be installed between the regions. A sump pump with failure alarms will manage water accumulation in the vault. Two 4'' conduits will connect the superstructure to the Building~92 manhole, providing paths for network and signal cables. A 24-strand fiber optic link will connect the superstructure to the second floor of the Counting House. Fire protection will be provided for both the vault and the superstructure. A lightning arrest system will also be installed on the superstructure to protect equipment and personnel.

This beamdump facility is being designed by the Facility Management Division at Jefferson Lab and the final drawings for the construction tender are under preparation. The realization of this vault is possible as early as 2026--27.


\section{Muons Beams and Physics}
%
\subsection{Muon Beams at the Beamdump Facility}
%
Major laboratories such as CERN, PSI, J-PARC, TRIUMF and Fermilab host muon beamlines. China and South Korea are planning muon sources. Surface muon beams are produced by the decay of positive pions stopped near the surface of a muon production target. The momentum is 27.4\,MeV$/c$, and the beam polarization is almost 100\,\%. Decay muon beams are generated from pion decays in flight, including positive and negative muons within a wide momentum range. Muon Spin Rotation/Relaxation/Resonance ($\mu$SR) techniques are tools for probing the local magnetic fields and electronic environments within materials. They are widely used to study superconductivity, magnetism, and phase transitions in complex materials. High-intensity muon beams enable searches for rare muon decays (e.g., $\mu \rightarrow e \gamma$ or $\mu \rightarrow eee$), which are sensitive probes of BSM physics. Stored high-energy muons facilitate precision measurements of the muon's anomalous magnetic moment ($g-2$), such as the experiments at Fermilab and J-PARC. Additionally, muon beams can explore dark sector interactions, such as hypothetical dark photons or light dark matter candidates produced via muon interactions. Experiments like Mu2e at Fermilab aim to detect the coherent conversion of a muon to an electron in the field of a nucleus, a SM-forbidden process but predicted in many new physics scenarios. An extensive international R\&D program is currently underway to develop and validate the required technologies for a multi-TeV muon collider and to explore viable technical solutions.

Muons are also a powerful probe for applications like tomography of large, dense objects, with uses in geological surveying, nuclear non-proliferation, and industrial inspection. However, the field is constrained by relying on cosmic rays characterized by a low and non-directional flux, hence requiring long exposure times, limiting muography to static targets.

\begin{figure}[htbp]
  \centering
  \includegraphics[width=0.32\textwidth]{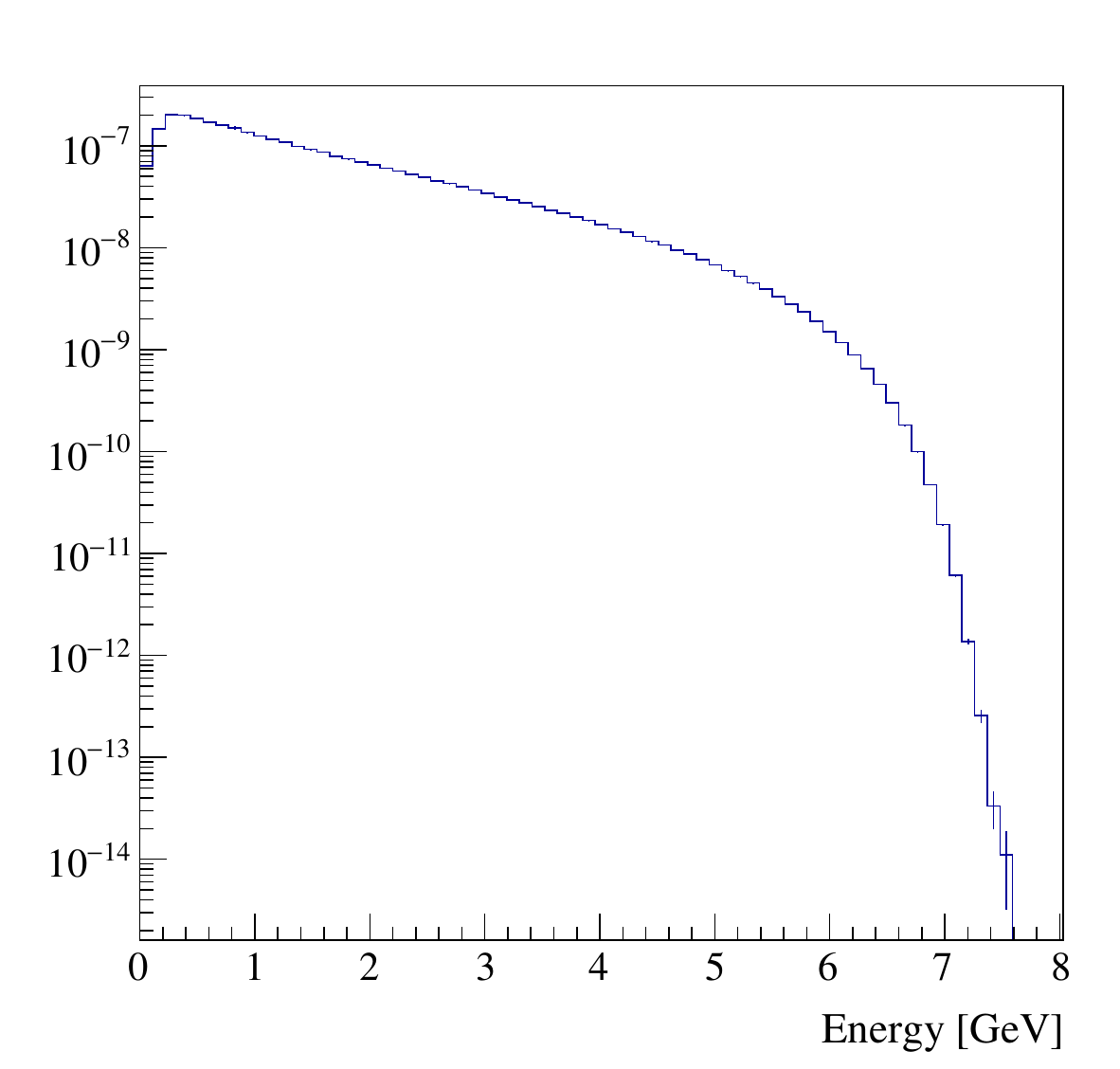}
  \includegraphics[width=0.32\textwidth]{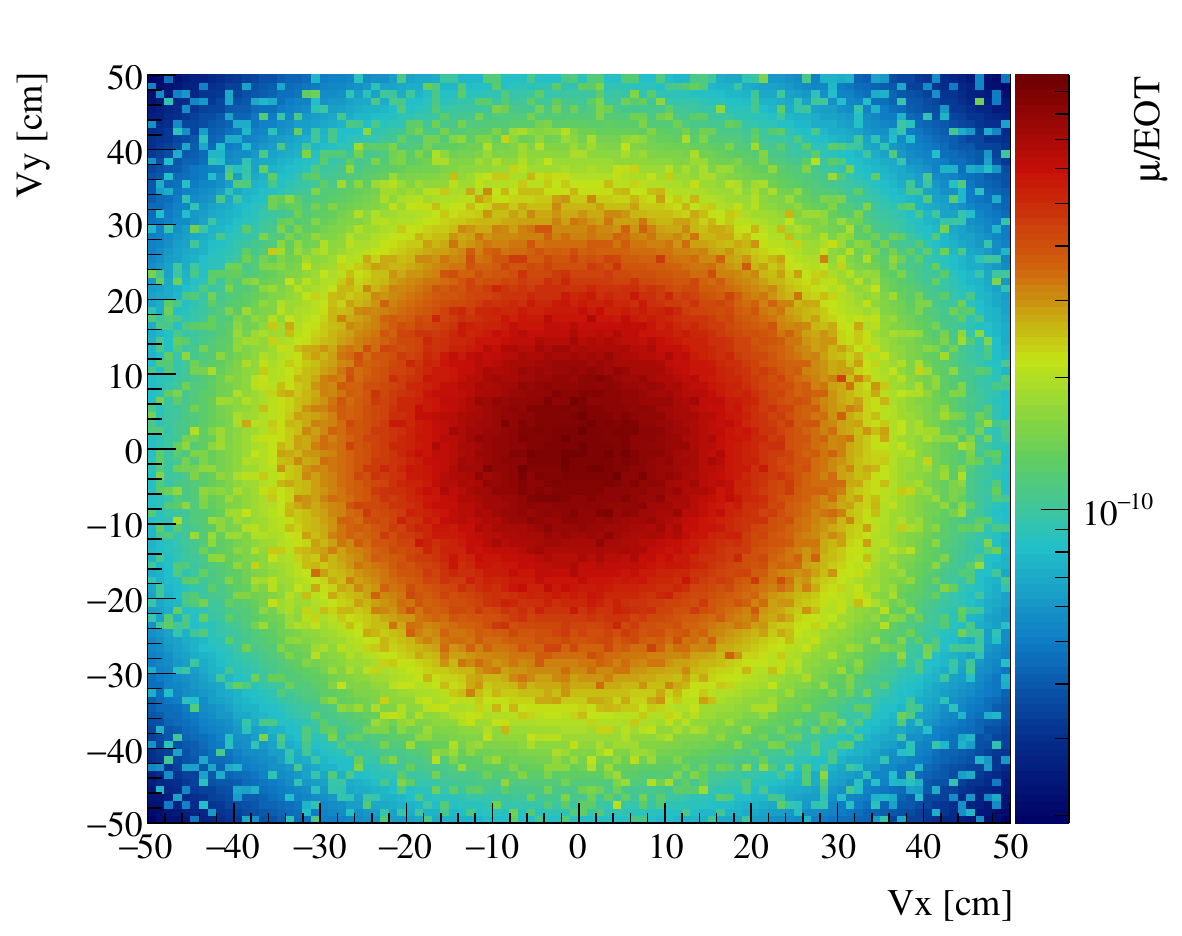}
  \includegraphics[width=0.32\textwidth]{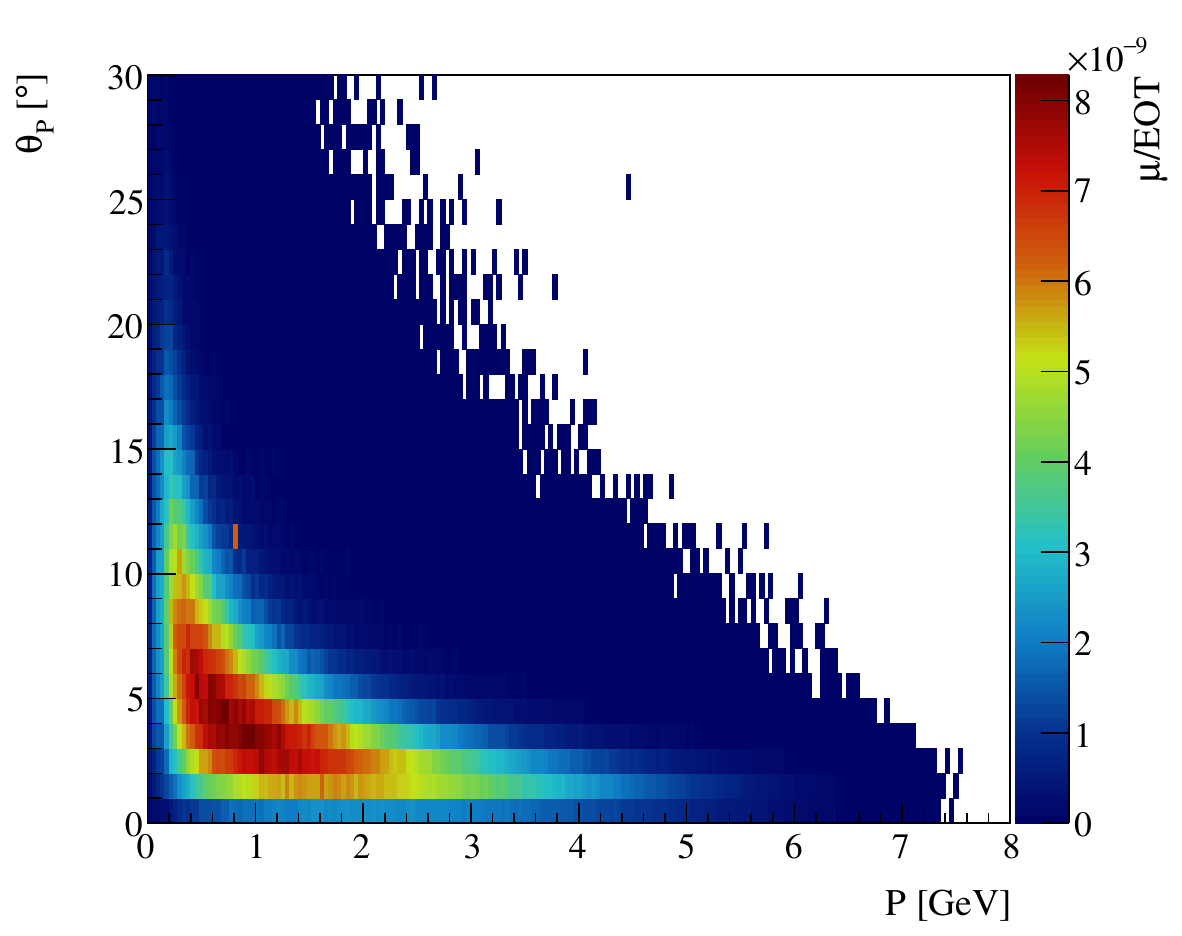}
  \caption{{\bf Left:} Muon energy spectrum for a primary 11-GeV electron beam. {\bf Center and right:} Spatial and angular distributions of the muon beam.}
  \label{fig:muonbeam}
\end{figure}

\begin{figure}[!t]
  \centering
  \includegraphics[width=0.495\linewidth]{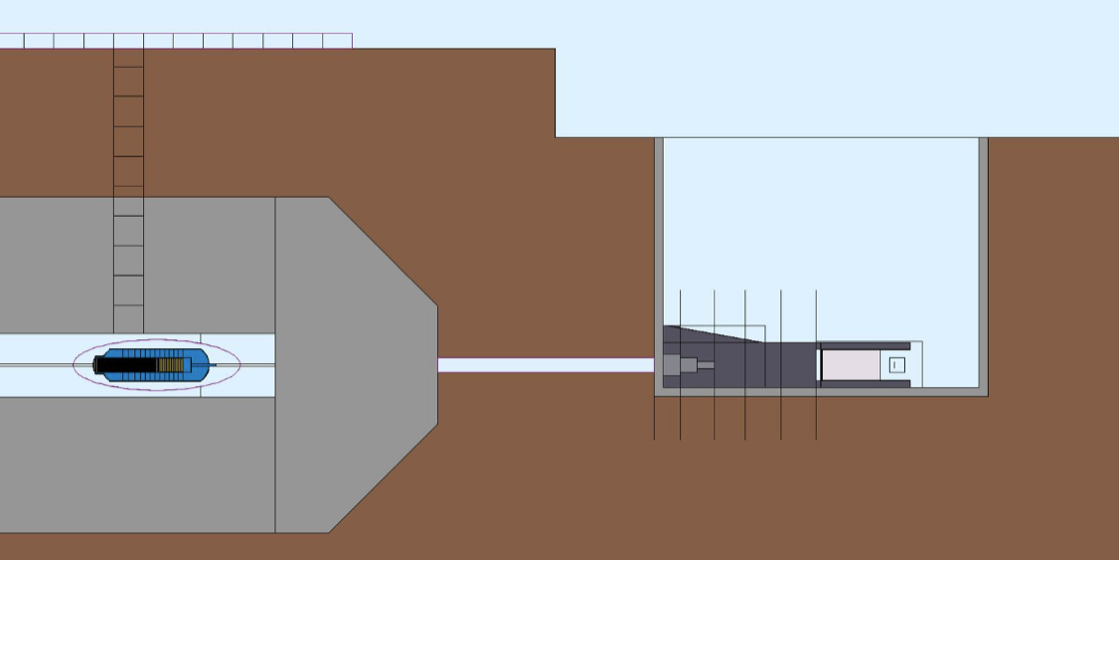}
  \includegraphics[width=0.495\linewidth]{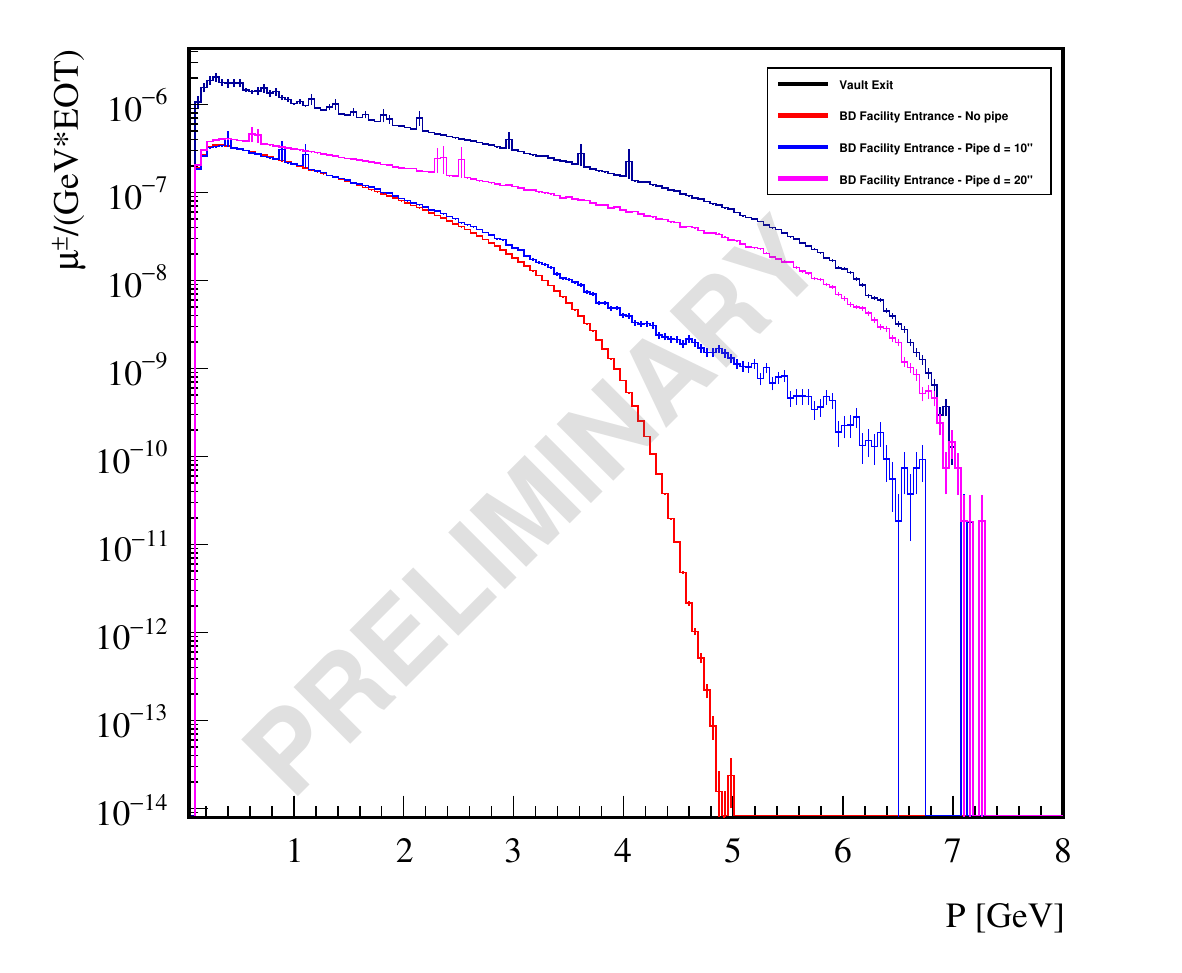}
  \caption{{\bf Left:} Geometry of the setup including a horizontal pipe connecting the vault exit to the entrance of the beamdump facility. {\bf Right:} Muon energy spectra at the exit of the beamdump vault and at the entrance of the beamdump facility with and without an air-filled 10"- or 20"-diameter pipe.}
  \label{fig:muon-pipes}
\end{figure}

The muon beams at Jefferson Lab originating from the beamdumps are unique, as they include both charge states with a bremsstrahlung-like energy spectrum extending up to 6\,GeV and high degrees of polarization at the higher energies. Pair-produced muons may have both left and right helicity for the muon of a given charge as opposed to muons from pion decay. At the exit of the Hall-A beamdump concrete vault, located 5.5\,m downstream, the flux through a plane of 1\,m$^2$ is up to $\sim 2.5\times 10^{-6}\,\mu/\mathrm{EOT}$, corresponding to an intensity of $8\times 10^8\,\mu/\mathrm{s}$ for an electron beam current of 50\,$\mu$A~\cite{Battaglieri:2023gvd}. At this position, the muon beam has a Gaussian 1-$\sigma$ width of 25\,cm. Figure~\ref{fig:muonbeam} shows the muon energy spectrum and the spatial and angular distributions. At the entrance of the beamdump facility the muon beam retains a similar energy spectrum, with a maximum at $\sim$ 4\,GeV, a total flux of $9\times 10^7\,\mu/\mathrm{s}$, and a 1-$\sigma$ width of 35\,cm. Preliminary simulations indicate that a 10''- (20''-)diameter horizontal pipe---connecting the exit of the concrete vault to the beamdump facility entrance and filled with air---would result in a flux of $1.8\times 10^{8}\,\mu/\mathrm{s}$ ($2.8\times 10^{8}\,\mu/\mathrm{s}$), a maximum energy of $6$\,GeV ($7$\,GeV). Figure~\ref{fig:muon-pipes}\,(left) shows the experimental set-up. Figure~\ref{fig:muon-pipes}\,(right) gives a comparison of muon energy spectra at the exit of the vault and at the entrance of beamdump facility with dirt (no pipe installed) and with 10''- and 20''-diameter air-filled pipes.

\subsection{Muon Beamlines at the Beamdump Facility}
%
Extending the beamdump facility upstream would make it possible to install a focusing system to reduce the muon beam size~\cite{Benesch:TN25-073}. 
\begin{figure}[ht]
  \centering
  \includegraphics[width=0.96\textwidth]{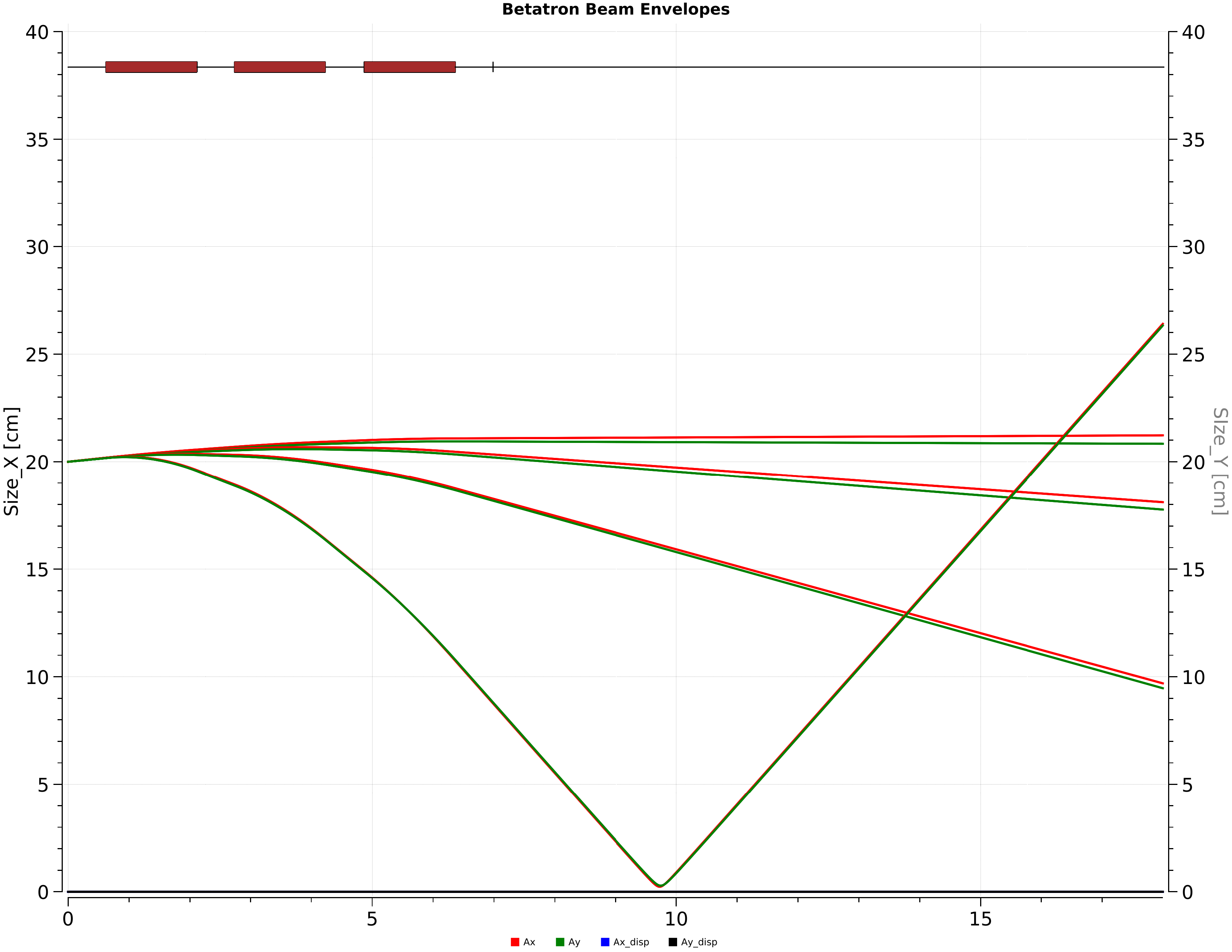}
  \caption{Design for muon beamline with
  focusing due to three 3\,T solenoids in persistent mode. The four pairs of lines, starting from the top, represent 8-GeV, 6-GeV, 4-GeV and 2-GeV muons. The exterior of the underground vault is represented by the tick mark at 7\,m at the top of the figure.}
  \label{fig:solenoid}
\end{figure}

\paragraph{Superconducting Solenoid Focusing}
Siemens sells a 7\,T whole body MRI scanner with self-contained cooling system. General Electric Medical Systems produces MRI systems up to 3\,T. 
Figure~\ref{fig:solenoid} suggests that trying to buy three 3-T 
units could be a viable solution. Correction coils are not needed for this purpose so one could also attempt to purchase units custom manufactured without them. In Fig.~\ref{fig:solenoid} the solenoidal field is 150\,cm and the cryostats are 200\,cm in length. Cryostat bore $\sim$ 90\,cm. These assumptions allow only 37.5\,cm each to concrete from the front of the first cryostat and the end of the last cryostat and 12.5\,cm between the cryostats. If the magnets were cooled with GM or pulse-tube refrigerators, not liquid He, they could be placed in an underground vault with only electrical service. Personnel access will be required on each side of the three-solenoid sequence as 37.5\,cm is insufficient per life safety code. If only two cryostats fit in the 7\,m available, focusing will be substantially reduced.

\paragraph{Quadrupole Focusing}

\begin{figure}[ht]
  \centering
  \includegraphics[width=0.48\textwidth]{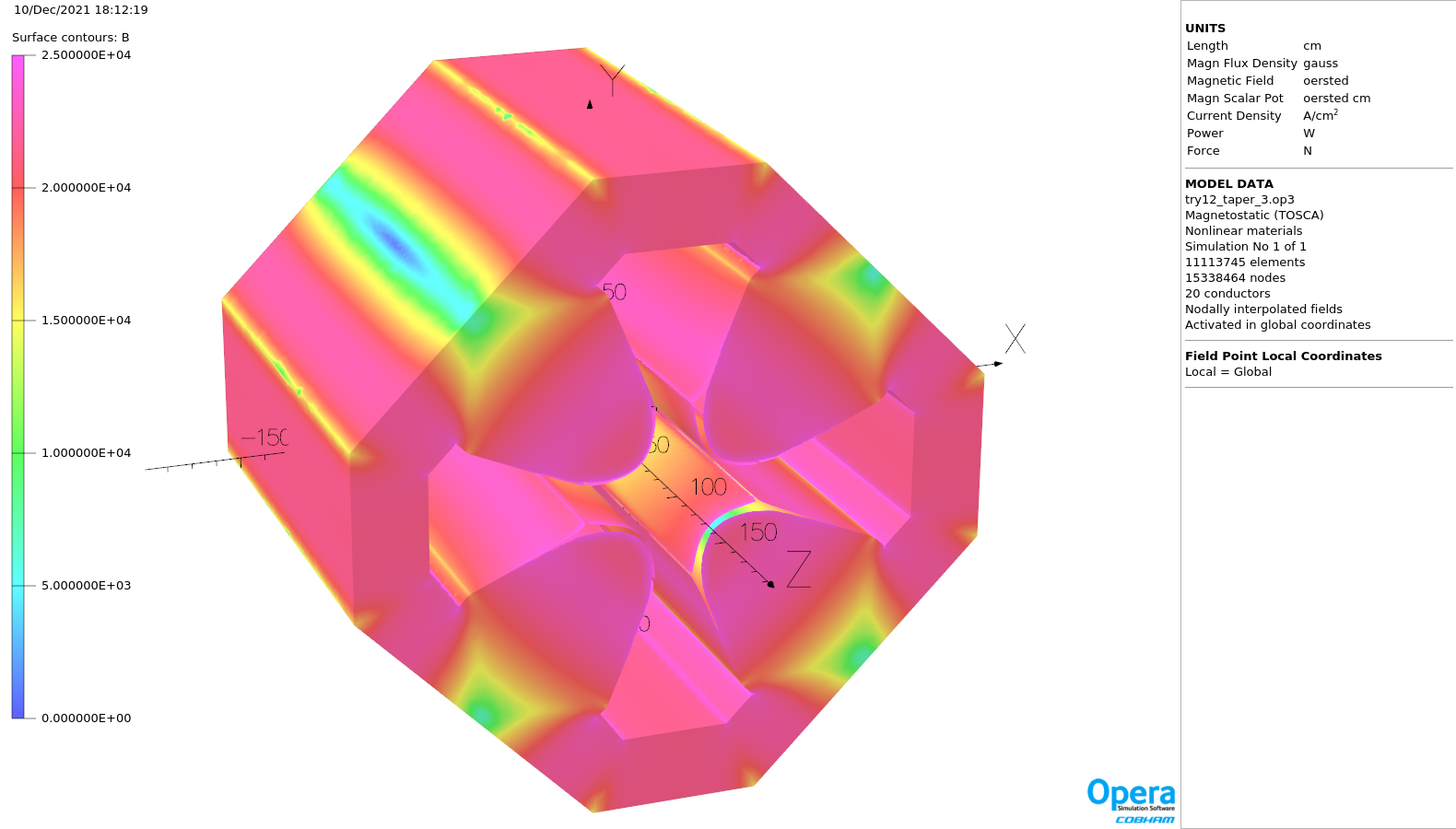}
  \includegraphics[width=0.48\textwidth]{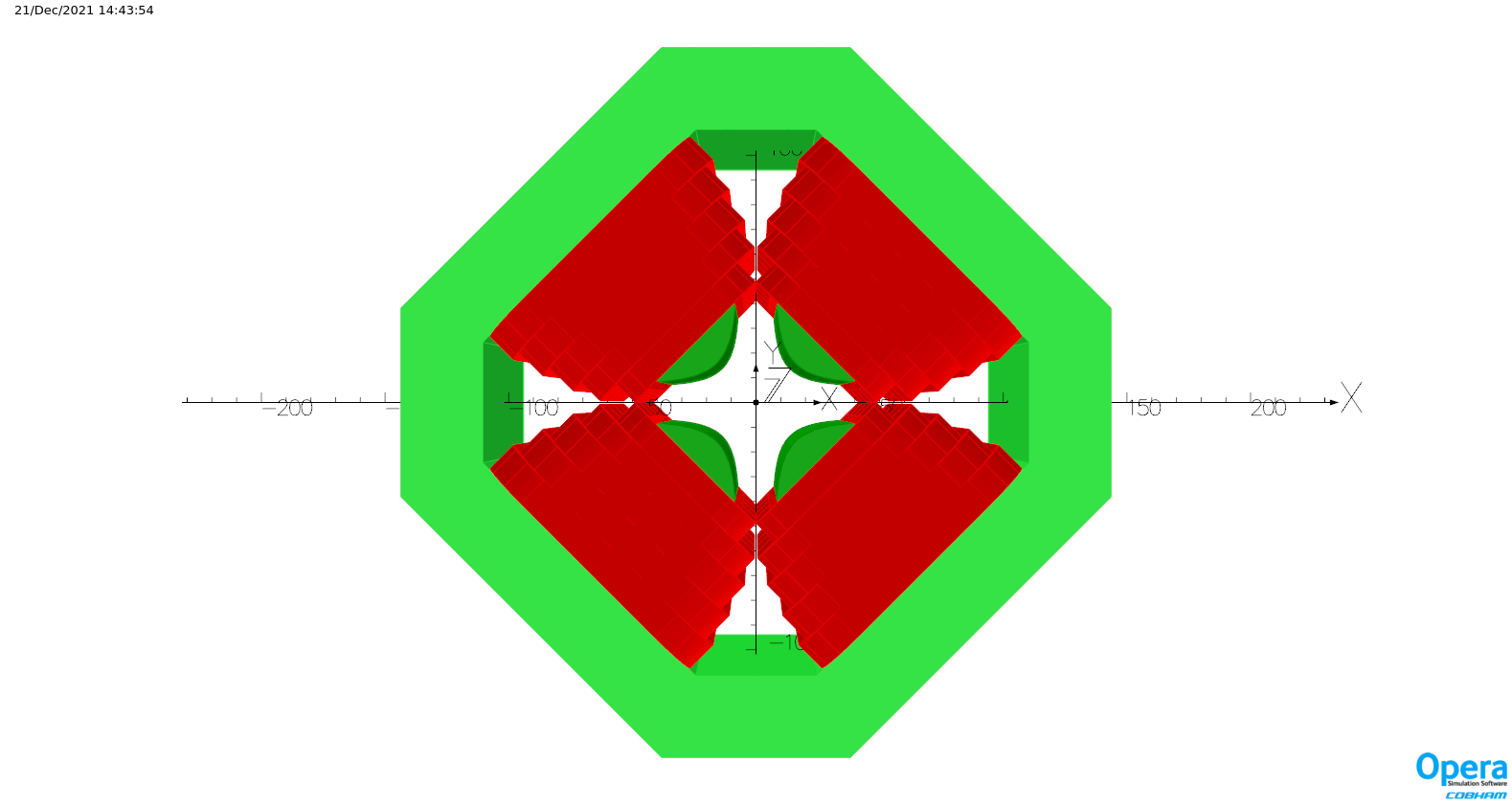}
  \caption{Quadrupole design for a muon beamline. {\bf Left:} Field on the surface for a current density of 300\,A$/$cm$^2$, pink corresponds to 2.5\,T. {\bf Right:} View in beam direction showing the 416\,mm aperture.}
  \label{fig:quad}
\end{figure}

A 104\,mm inner diameter by 350\,mm long quadrupole design 
was scaled to 416\,mm inner diameter by 2000\,mm length. 
Current densities from 25 to 400\,A$/$cm$^2$ were modeled: 275\,A$/$cm$^2$ proved a practical limit given low carbon steel saturation and the 20\,bar upper bound of cooling water pressure at Jefferson Lab.  At this current density the surface field on the poles approaches 2.5\,T. Figure~\ref{fig:quad} shows the design. Each quadrupole requires a 500\,kW power supply and 14\,l/s cooling water flow for 42\,K temperature rise.

Using this quadrupole a four quad FODO line was modeled. For 2\,GeV muons it has peak radius 400\,mm so the muon beam will be clipped but the muons remaining can be brought to a 200\,mm radius focus.  6\,GeV muons cannot be brought to a focus and peak radius in the line is 600\,mm so more will be clipped.  

\begin{figure}[ht]
  \centering
  \includegraphics[width=0.48\textwidth]{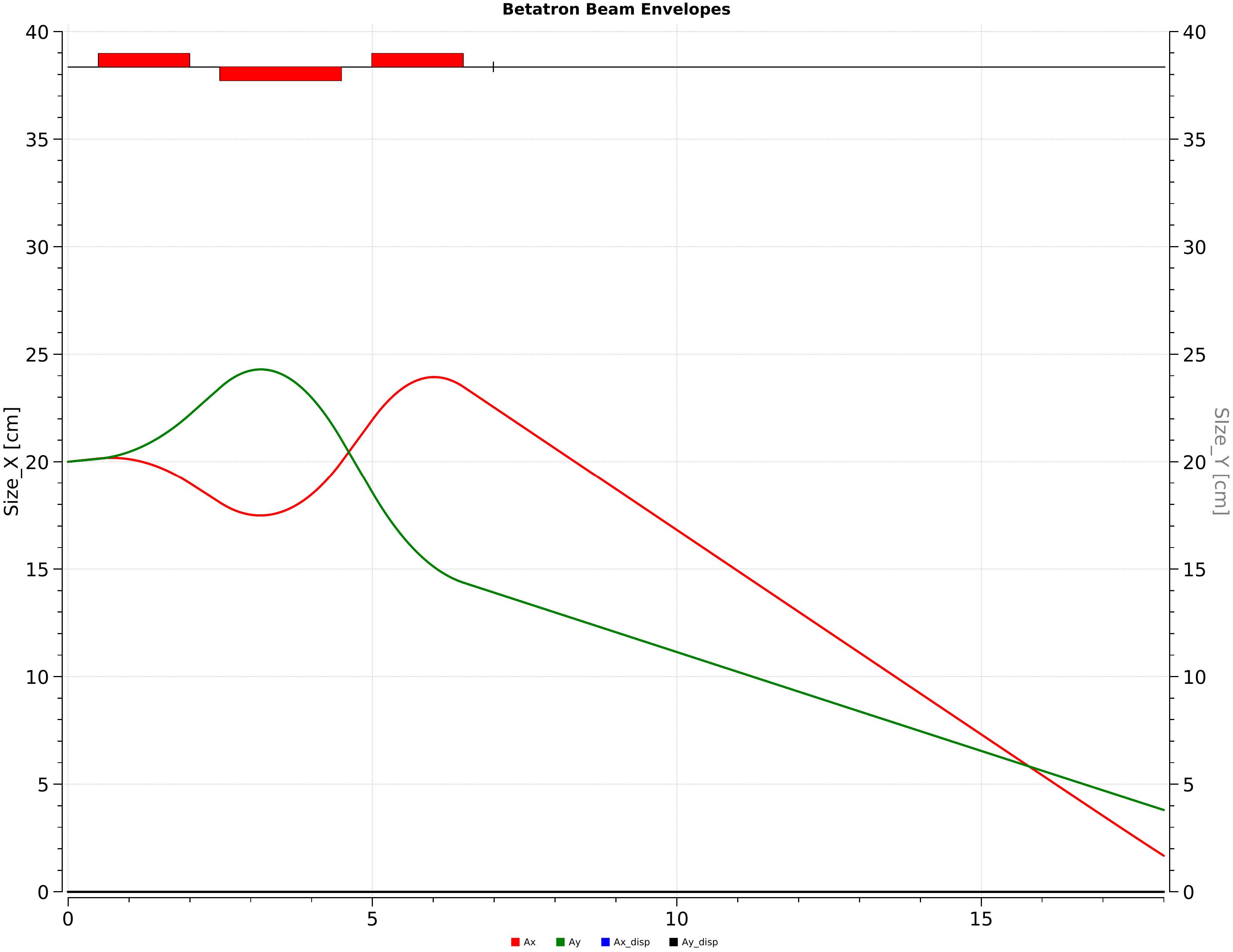}
  \includegraphics[width=0.48\textwidth]{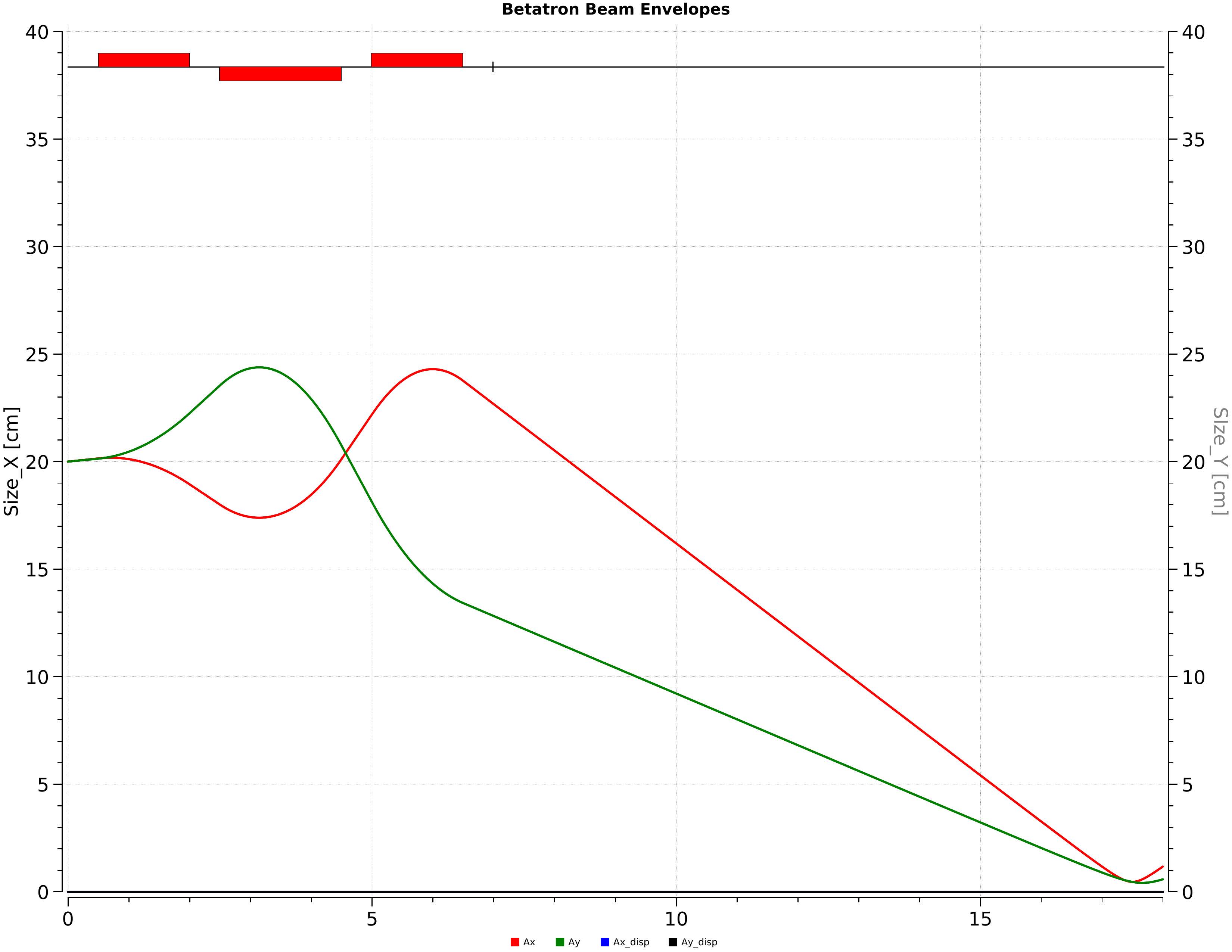}
  \caption{Design for a focusing muon beamline using a triplet of quadrupoles. {\bf Left:} 1-$\sigma$ size for 2-GeV muons. {\bf Right:} 1-$\sigma$ size for 6-GeV muons.}
  \label{fig:quadfocusing}
\end{figure}

Triplet focusing of the muons, assuming that half or more are discarded, also proves possible with the
416\,mm inner diameter quadrupoles. Lengths shown in the Fig.~\ref{fig:quadfocusing} are 150\,cm, 200\,cm and
150\.cm. These lengths could be shortened to 120\,cm,
160\,cm and 120\,cm. Building all three to 165.1\,cm (65'') or 167.6\,cm (66'') would reduce cost and
provide a few percent headroom for the middle magnet. 
Five-sixths is a reasonable multiplier for the power and cooling water requirements  if the 66'' length is chosen. Since these are resistive quadrupoles one can adjust the focusing to achieve roughly the same beam profile at the end of the vault for each momentum modeled. 
A beam pipe starts at $z = 7$\,m: If a 12” tube with 1” wall is used, this is 25.4\,cm diameter. With 16'' tube and 1'' wall, this is 35.6\,cm diameter and with 18'' tube and 1'' wall, this is 40.6\,cm diameter. The quadrupole inner diameter is 41.6 cm.


\paragraph{Other Considerations}
It is possible to focus the muons exiting the Hall A beamdump in the seven meters available between
the concrete enclosures of the dump and the vault. Use of 3-T superconducting solenoids may have
higher capital cost and would have much lower operating cost. Use of resistive quadrupoles would
have much higher operating cost but provide somewhat better momentum selection. Capital cost might
be similar because of the quadrupole size, required power supplies and a significant upgrade of the
Jefferson Lab's cooling water system. 
Moving the vault downstream one meter is advantageous if focusing is desired: both solenoids and quadrupoles are closer together than desirable for installation and maintenance. The quadrupoles with 167.6\,cm long steel have an overall length of 198\,cm including coils so the extra meter is imperative. A few\,cm of detectors might also be mounted between the magnets if this were done. Putting the vault as close to the road as possible would be good.

If the vault is doubled in area, to 5\,m by 12\,m, this would allow the steel shielding or either of the two muon focusing solutions to be installed inside. It should be placed to allow personnel access to either side of the beam line so equipment can be installed and maintained. The spiral staircase suggested at the workshop should be incorporated as well and could double as a forced ventilation chase if that is required given number of personnel allowed in the vault. The quadrupoles are $\sim$ 36 tonnes each, when estimated as 50\% dense 2.6\,m diameter by 1.7\,m long steel cylinder, so floor capacity should be at least ten tonnes per square meter. The 3-T superconducting MRI magnets would be approximately a tenth of this.

\subsection{Comparison with Existing and Planned Muon Facilities}

\begin{figure}[htbp]
  \centering
  \includegraphics[width=0.9\textwidth]{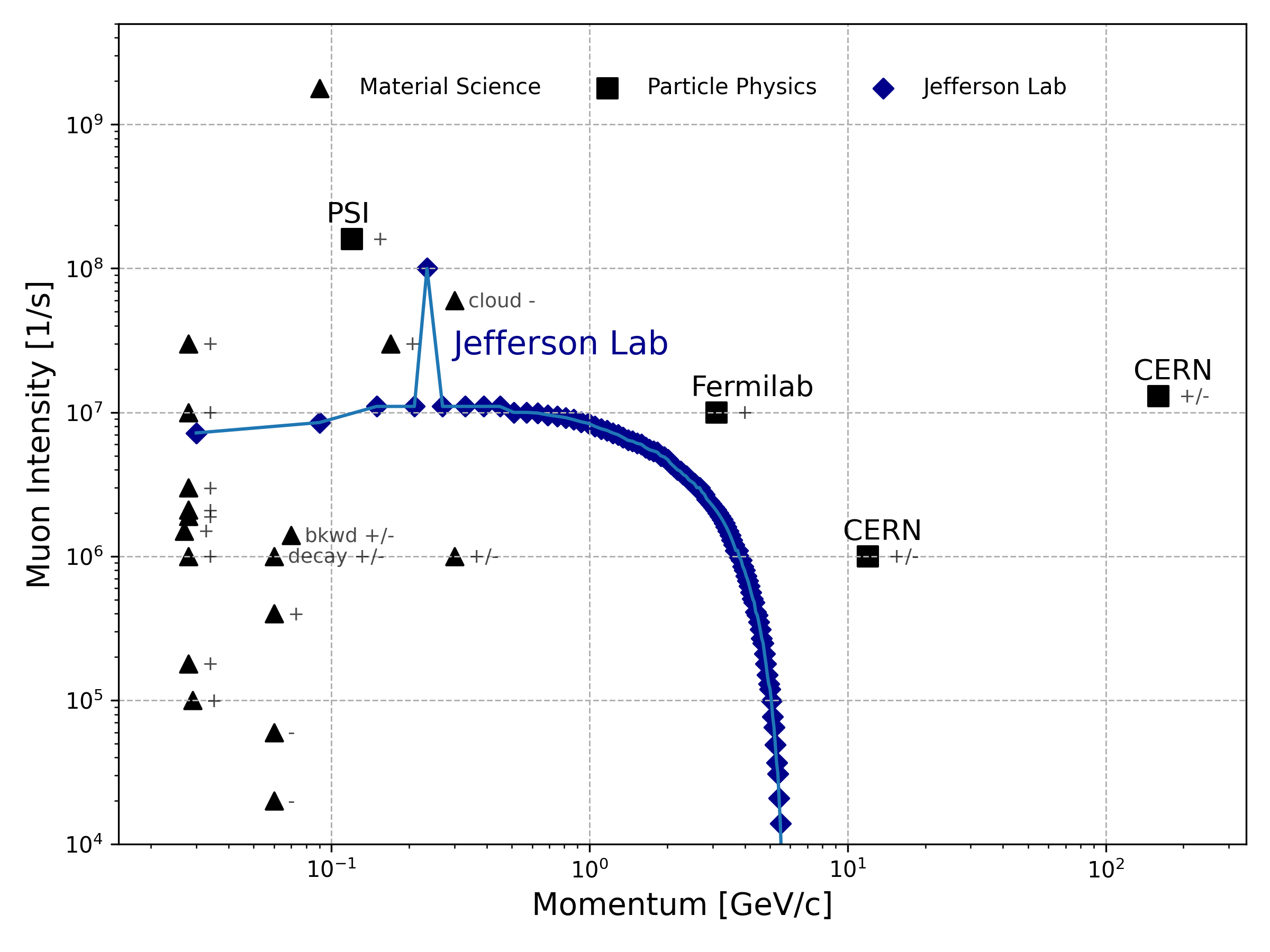}
  \caption{Comparison of muon beam intensities at different facilities. The listed intensity is indicative for the respective facility.}
  \label{fig:comparison}
\end{figure}

\renewcommand{\arraystretch}{1.4}
\begin{table}[htb]
\small
  \centering
  \caption{Comparison of different muon beam facilities. While the momentum range and beam type can refer to different beamlines and/or operation modes, the listed intensity is indicative for the respective facility.}
  \label{tab:facilities}
  \begin{tabular}{@{}p{4.2cm}cccl@{}}
  \toprule
 \textbf{Beamlines} & \textbf{Momentum} & \textbf{Intensity} & \textbf{Type} & \\ 
  & \textbf{\boldmath (GeV$/c$)} & \textbf{($\mu/$s)} \\
  \midrule
  CERN East Area \newline \footnotesize T9, T10 & 0.1 -- 15 & $2 \times 10^3$ & $\mu^\pm$ \\
  CERN North Area \newline \footnotesize M2 & 50 -- 220 & $5 \times 10^7$ & $\mu^\pm$ \\
  Fermilab (USA) \cite{Muong-2:2023cdq} \newline \footnotesize MTest, g-2 beamline & 3.1 & $1 \times 10^7$ & $\mu^+$ \\
  ISIS RIKEN-RAL (UK) \cite{hillier:2019:RIKEN} \newline \footnotesize Port 1/2, Port 3/4 & 0.02 -- 0.12 & $5.9 \times 10^5$ & surf., $\mu^\pm$ \\
  J-PARC\,MLF (Japan) \cite{Miyake:2009,Kawamura:2021} \newline \footnotesize D, U, S, H & 0.002 -- 0.12 & $1 \times 10^8$ & surf., $\mu^\pm$, cloud $\mu^-$ \\
  PSI (Switzerland) \cite{PSI} \newline \footnotesize PiE5, PiE3, PiM3, MuE4, MuE1 & 0.01 -- 0.12 & $1.6 \times 10^8$ & surf., $\mu^\pm$, cloud $\mu^-$  \\
  RCNP (Japan) \cite{MuSIC} \newline \footnotesize DC Beam Line – M1 & 0.03 -- 0.08 & $6 \times 10^4$ & $\mu^\pm$ \\
  TRIUMF (Canada) \cite{TRIUMF-Muons,TRIUMF-muSR} \newline \footnotesize M13, M15, M20 & 0.02 -- 0.30 & $1.4 \times 10^6$ & surf. \\
  \midrule
  Jefferson Lab \cite{Battaglieri:2023gvd} \newline \footnotesize Beamdump Facility & $<$ 6 & $1 \times 10^8$ & pol. $\mu^\pm$ \\
  \bottomrule
  \end{tabular}
\end{table}

Figure~\ref{fig:comparison} shows a comparison between the muon beam intensity at the beamdump facility and that of existing beams at other facilities worldwide. Table~\ref{tab:facilities} lists the momentum ranges, intensities, and types of muons beams. 
\paragraph{Muon Beamlines at CERN}
%
CERN provides a wide range of secondary beams for detector R\&D and fixed-target experiments. In the North Area, originating from the 400\,GeV$/c$ primary proton beam of the SPS, the M2 beamline in EHN2 offers tunable muon momenta between 50\,GeV$/c$ and 220\,GeV$/c$ with $\ge 3.7$\% momentum spread with a maximum intensity of $5 \times 10^7\,\mu/$s with future upgrades post Long Shutdown 3 aiming at $1 \times 10^8\,\mu/$s.  The secondary beam lines in EHN1 in the North Area can also provide high energy, wide momentum spread muon beams with a maximum intensity of $10^4\,\mu/$s. Additionally, the East Area secondary beamlines, T9 and T10, offer muons beams in the range of 0.1\,GeV$/c$ to 15\,GeV$/c$, at a maximum intensity of $2 \times 10^3\,\mu/$s with a wide momentum spread ($\sim 10 - 15\%$). Together, the CERN North and East Area facilities constitute one of the most versatile infrastructures for secondary and tertiary beams worldwide, serving over 200 user teams annually.

\paragraph{Muon Beamlines at Fermilab}
%
Fermilab has a broad range of accelerators. Pions are collected from the Booster synchrotron into the Recycler ring, which are then sent into the muon beam storage rings (E989 and E821) to be used for the Mu2e and $g-2$ experiments. The muon beam is set at a ``magic momentum'' of 3.1\,GeV$/c$. The use of electrostatic quadrupoles doesn't affect the muon spin, which allows for the precise measurement of the muon's anomalous magnetic dipole moment~\cite{Muong-2:2023cdq}.

\paragraph{Muon Beamlines at ISIS RIKEN-RAL}
%
The RIKEN-RAL muon facility at the ISIS Neutron and Muon Source~\cite{hillier:2019:RIKEN} at the Rutherford Appleton Laboratory (RAL) in the UK originated from a collaboration with RIKEN in Japan. A proton beam from the ISIS accelerator (800\,MeV, typically 200\,$\mu$A) is producing positive ($4 \times 10^5\,\mu^+/$s at 60\,MeV$/c$) and negative ($7 \times 10^4\,\mu^-/$s at 60\,MeV$/c$) muon beams with a momentum selected between 17 and 120\,MeV$/c$. A surface muon beam at 27\,MeV$/c$ with $1.5 \times 10^6/$s intensity is also available. The facility supports both fundamental and applied research. Among the research activities at ISIS, 
%
the FAMU experiment~\cite{pizzolotto:2020:FAMU}  
is a precision measurement of the ground state hyperfine splitting in muonic hydrogen with an accuracy of $\sim 10^{-5}$,  
enabling the evaluation of the proton Zemach radius with unprecedented accuracy lower than 1\%, along with testing QED corrections. 
The muonic hydrogen atoms are produced with the RIKEN-RAL Port1 pulsed negative muon beam, with an average pulse rate of 40\,Hz, and set to a momentum of 55\,MeV$/c$, and a monitored intensity of $\mathcal{O}(10^4)$ muons$/$s~\cite{rossini:2024:hodoRAL}.
Muons are injected in a target 
containing a multi-pass optical cavity. 
When the $\mu$H atoms are thermalized, 
medium infrared laser light 
is injected in the target. 
The marker for the transition is an excess of muonic oxygen characteristic X-rays 
following the laser injection. The experiment is currently running. 
In addition, 
%
a prominent activity 
at ISIS is the elemental and isotopic non-destructive characterization of samples using muonic atom X-ray emission analysis ($\mu$-XES). This technique is based on the detection and quantification of the X- and $\gamma$-rays emitted by a sample when irradiated with negative muons. The beam momentum sets the depth of penetration of muons in the sample, allowing for depth-dependent measurements. 
Muon imaging and isotopic measurements are also under development. At RIKEN-RAL, one of the irradiation ports is equipped with a $\mu$SR spectrometer capable of using surface and decay muons of either polarity, whereas at ISIS three additional $\mu$SR instruments operate exclusively with surface muons. These facilities are primarily dedicated to condensed matter research.
%


\paragraph{Muon Beamlines at J-PARC MLF}
At the J-PARC MLF facility, high-intensity muon beams are produced by a 3-GeV proton beam (beam power 1~MW, repetition rate 25~Hz) impinging on a 20~mm thick graphite target. Pions generated by nuclear reactions of the protons in the target decay producing muons and neutrinos. Two different modes are commonly used to extract muons as secondary beams: surface muons and decay muons. Currently, four muon beamlines (the D-, U-, S-, and H-lines) are expected to extend from the target to the experimental halls. These beamlines are designed to provide high-flux muon beams with different properties to meet the requirements of a variety of muon experiments~\cite{Kawamura:2021jps}.

The D-line provides surface muons with an intensity of 10$^{7}$/s as well as decay positive$/$negative muons with momenta of up to 120\,MeV$/c$ with a typical intensity of $10^{6} - 10^{7}/$s at 60\,MeV$/c$. The experimental area served by the D-line hosts a a $\mu$SR spectrometer and Ge detector for muonic X-ray analysis. 
The U-line, under commissioning, is designed to produce an ultra-slow muon beam (kinetic energy from a few eV to 30\,keV.)~\cite{Miyake:2014jps} .
A $\mu$SR spectrometer is installed in the experimental area for ``nm-$\mu$SR'' for thin  sample irradiation and surface analysis, while, for  thicker samples,
moreover a compact AVF cyclotron re-accelerates ultra-slow muons up to 5\,MeV. 
%
%
In the S-line 
surface muon beams are used for
$\mu$SR experiments~\cite{Kojima:2018jps} and laser spectroscopy of the muonium 1S--2S energy splitting.
%
%
The H-line is a high-intensity muon beamline for general use~\cite{Yamazaki:2023epj}.
Surface muon beams with intensities of up to $10^8/$s and cloud muon beams up to 120\,MeV$/c$ are available.
Precision measurements of the hyperfine structure of muonium 
~\cite{Shimomura:2011aip}
and a search for $\mu$-e conversion~\cite{Natori:2014npb} are underway.
Re-accelerated muons up to 4\,MeV will become available in 2027.
%
The H-line is planned to be extended to develop a novel low-emittance muon beamline~\cite{Abe:2019ptep}. 




\paragraph{Muon Beamlines at LBNL}
Laser-plasma accelerators, which use PW-class lasers to generate 10s of GV$/$m accelerating gradients,
produce multi-GeV electron beams in plasma targets that are 10s of cm in length.
%
In a recent experiment at the BELLA Center at LBNL,
muon beams were generated using an electron beam with energies extending up to 8 GeV driven by a laser-plasma accelerators.
The electron beam was stopped in the beam dump,
and the resulting muons were detected behind a 90-cm concrete wall.
Muon production was unambiguously confirmed by capturing delayed signals in plastic scintillators,
corresponding to the decay of stopped muons.
%
Muon production in this environment is driven by two primary mechanisms:
(i) a forward-directed, high-energy muon beam from the Bethe--Heitler process in the main beam dump, and
(ii) a quasi-isotropic, lower-energy flux from meson decays.
%
At a typical repetition rate of $\sim 1$\,Hz, this flux is orders of magnitude more intense than the cosmic ray background for horizontal muography,
paving the way for compact, transportable systems for rapid imaging.



\paragraph{Muon Beamlines at PSI}
%
The Paul Scherrer Institute (PSI) in Switzerland is a leading center for neutron and muon science. Its 590-MeV proton accelerator, operating at currents up to 2.3\,mA, delivers a beam power of over 1.3\,MW. This beam is directed to two meson production targets (Target M and Target E), which produce pions and muons. After passing through these targets, the remaining proton beam is directed to the Swiss Spallation Neutron Source (SINQ) for neutron production. PSI operates seven secondary beamlines for muon and pion extraction. These beamlines serve both particle physics and material science experiments. Two beamlines are dedicated to particle physics, including $\pi$E5, a high-intensity pion and muon beamline with a momentum range of 10--120\,MeV$/c$. It can deliver up to $5 \times 10^8$\,muons$/$s at 120\,MeV$/c$. 
%
%

%
PSI is well known as a leading facility for searches for charged lepton flavor violation with muons~\cite{Signer:2021hxr}. 
%
%
%
These experiments and other high-precision experiments with muonic atoms and muonium provide tests of QED and low-energy QCD and benchmarks for nuclear physics and weak interactions. 
%
%
The installation of high-intensity muon beamlines is part of the ongoing IMPACT project~\cite{Kiselev:2023ret} that will allow to deliver roughly two orders of magnitude increased intensities to experiments. 
The beamlines $\pi$E3, $\pi$M3, and $\mu$E4 are used for $\mu$SR and other material science applications. These typically operate with surface muons at 28\,MeV$/c$, with different intensities in the $10^6 - 10^8\,\mu^+/$s range. The $\pi$E1 beamline provides both positive and negative muons with a wide momentum range of 10 -- 500\,MeV$/c$, depending on the optical mode used. It is one of the most versatile beamlines at PSI.

\paragraph{Muon Beamlines at RCNP}
%
MUon Science Innovative muon beam Channel (MuSIC) at Osaka University is a muon beam facility with a continuous time structure. The muon generation employs a pion production target made of a long graphite placed in a 3.5~T solenoid magnetic field, which produces muons more than 1000 times more efficiently than conventional methods. This technology allows MuSIC to reach a maximum momentum of 80\,MeV$/c$. For positive muons the typical beam intensity at 30\,MeV$/c$ is $1 \times 10^5/$s. For negative muons the momentum the typical beam intensity at a momentum of 60\,MeV$/c$ is $2 \times 10^4$. In addition to particle and nuclear experiments and muon non-destructive analysis, experiments are also being carried out using a muon spin rotation measurement device for condensed matter research installed at the exit of the beamline~\cite{MuSIC}.

\paragraph{Muon beamlines at TRIUMF} 
%
The Center for Molecular and Materials Science at TRIUMF serves four beam lines M15, M13, M20, and M9B for spin-polarized muons ($\mu^+$ or $\mu^-$), which are implanted in materials and characterized with $\mu$SR (muon spin rotation/relaxation/resonance) and $\beta$NMR (beta-detected nuclear magnetic resonance) techniques. M15 and M13 transport momenta between $\sim$ 20 to 30 or 40 MeV$/c$. M15 provides an intensity of $1.9 \times 10^6/$s, M13 an intensity of $1.8 \times 10^5/$s. M20 and M9B reach 200\,MeV$/c$ in momentum. M20 delivers a $\mu^+$ intensity of $2.1 \times 10^6$, M9B delivers a $\mu^-$ intensity of $1.4 \times 10^6$~\cite{TRIUMF-Muons,TRIUMF-muSR}.

\paragraph{Future Muon Beamlines at SHINE}
Possible muon beam at the Shanghai SHINE facility have recently been discussed~\cite{Liu:2025ejy}. For SHINE, also surface muons have been considered.

\subsection{Prospects for Muon Physics at the Beamdump Facility}
Several outstanding topics of physics can be addressed with muons: verification of $e-\mu-\tau$ universality, limits on lepton number violation and physics beyond the standard model; nuclear physics studies with muons, lepton-mass effects and QED corrections; 
muon sources as a testbed for the Muon Collider R\&D on muon generation and beam cooling; physics and recent progress in muon-facilitated nuclear fusion.
Given the huge progress made in the past years in the computation of QED corrections to low-energy scattering processes~\cite{Aliberti:2024fpq}, the bottleneck towards further improvements is now often due to the impact of non-perturbative hadronic effects. By turning the argument on its head, this can be exploited to obtain additional information on non-perturbative effects from precision experiments. Prime examples are the extraction of the hadronic vacuum polarization through muon--electron scattering by MUonE~\cite{MUonE:2016hru, Abbiendi:2677471} or learning about two-photon exchange contributions in lepton--proton scattering~\cite{Afanasev:2023gev}. Studying processes with external muons rather than electrons can offer advantages, since QED corrections tend to be smaller and, hence, non-perturbative effects are enhanced w.r.t.\ remaining uncertainties. 
Additional input from muon experiments with secondary beams  can be used to improve the knowledge of non-perturbative effects in QED calculations.

As an example for a nuclear physics topic, the process $\mu^\pm\,N\to\mu^\pm\,N\,e^+\,e^-$ could be studied. This process is one of the main backgrounds for the MUonE experiment, and experimental data can be used to test the theoretical description. In the most advanced calculation~\cite{Abbiendi:2024swt} this background process is described as the scattering of a muon in the external Coulomb field, using a form factor to describe the nuclear charge distribution.

\subsubsection{Elastic Muon--Proton Scattering}
There is significant interest from both theoretical and experimental communities in determining the size of the hard two-photon exchange (TPE) effect. The hard TPE can be accessed directly through experiment by measuring the cross section ratio $R_{\pm}^\text{exp.}=\frac{\sigma(\ell ^+p)}{\sigma(\ell^-p)} \approx 1 - 2 \delta_{2\gamma}$. Recently the TPE has been studied with $e^{\pm}$ beams~\cite{Rachek:2014fam,A1:2013fsc,PhysRevLett.118.092501}. The effect appears to be small, approximately 1\%, in the measured region. However, these experiments did not have a sufficient reach in the four-momentum transfer ($Q^2$) region where the hard TPE effect is predicted to be more pronounced. The TPE effect can also be studied with muons. Currently there are two experiments taking data with $\mu^{\pm}$ beams, COMPASS++/AMBER~\cite{Adams:2676885} and MUSE~\cite{Cline:2021ehf}, that aim to measure two-photon effects. Both experiments have limitations in their sensitivity to the TPE. MUSE measures at low $Q^2$ while COMPASS++/AMBER measures at $\varepsilon \approx 1$. 

\begin{figure}
  \includegraphics[width=0.48\textwidth]{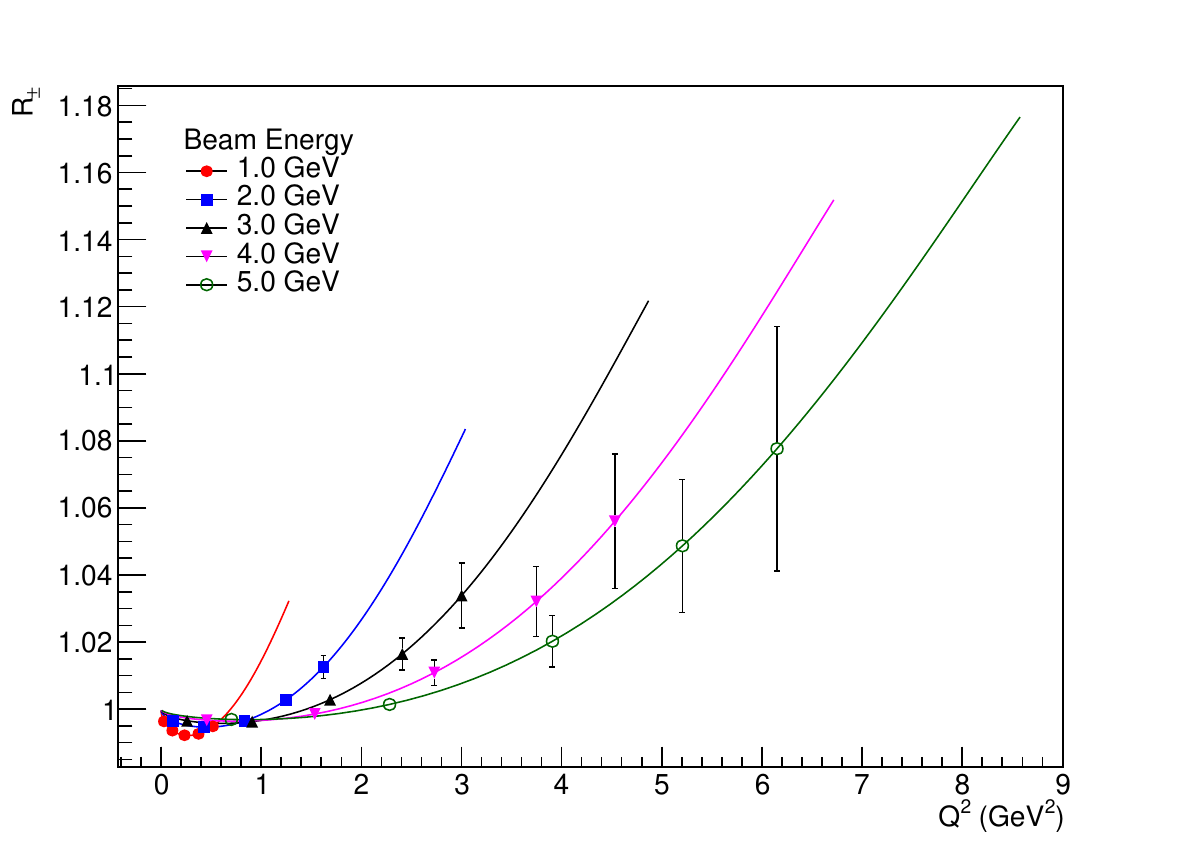}
  \includegraphics[width=0.48\textwidth]{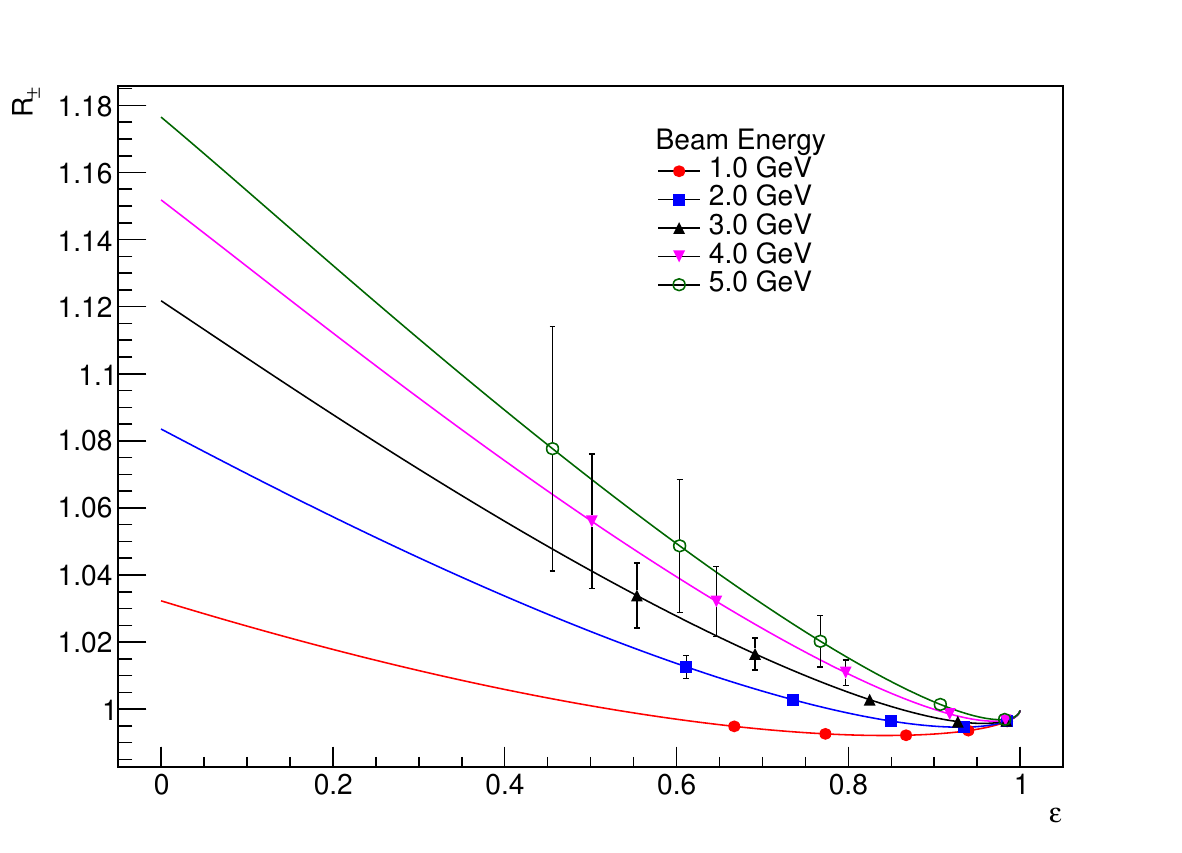}
   \caption{Size of the predicted TPE effect as a function of $\varepsilon$ and $Q^2$ for five muon energies. The points shown represent statistical uncertainties only assuming a one meter long liquid hydrogen target and a muon intensity of $1 \times 10^7$\,$\mu/$s. The uncertainties are calculated for one year of data taking, and detection at scattering angles $\theta = \{10, 20, 30, 40, 50 \}^{\circ}$ in $5^{\circ}$ wide bins in $\theta$.}
  \label{fig:TPE}
\end{figure}

A novel experimental design situated at a high-energy muon beamline could be constructed to take advantage of a wide range of $Q^2$ and $\varepsilon$ in order to probe the hard TPE in a region where it is predicted to be large. Jefferson Lab's muon beam can provide a unique opportunity to measure $R_{\pm}$. The size of the TPE effect using the parameterization from~\cite{A1:2013fsc} is shown in Fig.~\ref{fig:TPE}. It is worthwhile to note that all energies and both charge polarities are in the beam at once, potentially allowing for simultaneous measurement. However, in order to properly reconstruct scattering events, the initial particle momentum and species must be known. 

\paragraph{Targets and instrumentation required}
Given the complex beam properties, a series of incoming beamline detectors will be needed. Tracking detectors to measure the incoming beam trajectory will allow for a precise scattering vertex determination in conjunction with a scattered particle tracker. A large-area GEM detector would be well suited as an incoming tracker. For a scattered particle spectrometer it will be important to cover the largest possible solid angle. A magnetic spectrometer would be needed to identify elastic scattering events, and reject muon decay events. The hydrogen target requires a large scattering chamber. The MOLLER target is 125\,cm in length and the chamber is 2\,m in diameter.  

\paragraph{Required infrastructures}
A muon capture beamline will likely be required to maximize the incident beam flux on the target. Depending on spectrometer design, some excavation of the vault may be necessary to accommodate the spectrometer as it moves to backward angles.

The underground vault as proposed is likely sufficiently long to install it, but is narrow enough that installing a detector that has significant reach in scattering angle will be a challenge.

\subsubsection{Muon--Electron Scattering}
Muon-electron scattering of two pointlike fermions is a fundamental textbook process. At high muon energies of 160 GeV available to MUonE at CERN, this process is used to probe hadronic vacuum polarization, a higher-order radiative correction relevant for accurate calculations of the muon's $g-2$ value.

With the availability of a secondary muon beam at Jefferson Lab, using a diffuse target, tracking and timing before and after the scattering, as well as particle ID and calorimetry of the outgoing muon and electron, would allow to measure this fundamental SM process. With available muon energies up to a few GeV, such a muon-electron scattering experiment would overlap with the onset of the phase space covered by MUonE at CERN using 160 GeV muons, and it would be suited to probe fundamental fermionic cross sections and radiative effects with high precision.

\paragraph{Targets and instrumentation required}
The detectors needed for this measurement are similar to the ones proposed for the TPE physics: 
beamline detectors for determine incoming trajectories and particle momenta, along with scattered particle spectrometers instrumented with GEM tracking detectors and trigger hodoscopes.

\paragraph{Required infrastructures}
 The experiment would benefit from a  muon capture beamline. 

\subsubsection{Muon--Nucleus Scattering}

\begin{figure}[htbp]
  \centering
  \includegraphics[width=0.6\textwidth]{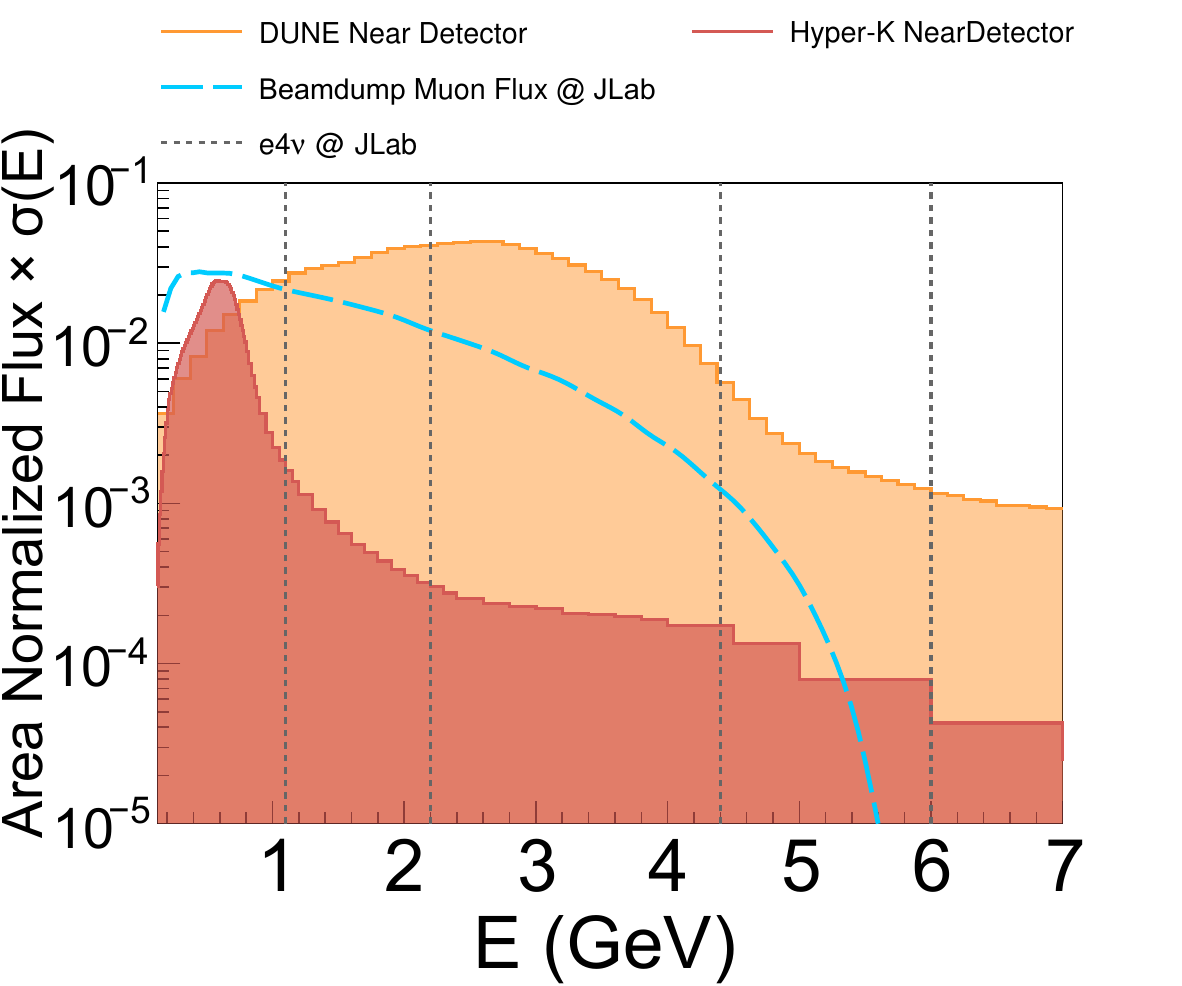}
  \caption{Comparison of flux-weighted cross sections relevant for neutrino experiments. The corresponding fluxes are normalized to the same area for ease of comparison. Shown are the spectra for the DUNE Near Detector (orange) and Hyper-K Near Detector (red), compared to the expected muon fluxes at the beamdump facility (dashed light blue). The vertical dashed lines indicate the energy coverage of the e4$\nu$ program at Jefferson Lab.}
  \label{fig:neutrino_flux}
\end{figure}

Jefferson Lab’s e4$\nu$ program leverages 1--6\,GeV electron-scattering data from CLAS to benchmark and constrain neutrino event generators~\cite{Papadopoulou_2021,Khachatryan2021}. Complementary to the e4$\nu$ program, muons can be exploited to probe the nuclear response in lepton--nucleus interactions, with the added advantage of smaller radiative effects compared to electrons~\cite{RadCorrections}. The muon beam at the beamdump facility offers a unique opportunity for high-precision lepton--nucleus measurements relevant for neutrino physics. With a bremsstrahlung-like energy spectrum extending up to 6\,GeV for the 11-GeV electron beams it enables detailed studies of lepton--nucleus cross sections. Furthermore, if the facility could provide the capability to tag muon energies with percent-level accuracy, this would allow the e4$\nu$ program to be extended in the sub-GeV region beyond the its current reach, see Fig.~\ref{fig:neutrino_flux}, and would make it interesting also for the Hyper-K experiment. In addition, one could quantify differences between muon-- and electron--nucleus scattering cross-sections at low energies. By combining the e4$\nu$ and the beamdump muon physics program with direct CEvNS measurements, the facility enables powerful cross-checks that help reduce systematic uncertainties for neutrino experiments.

\subsubsection{Muon Tomography}
%
Practical muography for imaging requires detecting each muon's position and direction before and after the object of interest. The trajectory is then reconstructed to build up a density map~\cite{schultz_image_2004,vanini_muography_2018}.
The number of muons required to image a meter-scale object is $10^6-10^9$, depending on the desired resolution. The high-current electron beam at Jefferson Lab is expected to deliver this number of muons on a timescale of minutes. The large muon flux available at the beamdump facility would not only enable rapid, high-resolution imaging far beyond the capabilities of cosmic rays or current laser-plasma accelerators, but also open up new opportunities for other high-statistics muon experiments.

\paragraph{Targets and instrumentation required}
To measure the muon trajectories, position-sensitive detectors such as silicon-based trackers or scintillator hodoscopes would be placed before and after the imaging sample. Ideal detectors should have a fast recovery time, on the order of tens of nanoseconds, to mitigate event pile-up. Suitable technologies include detectors based on fast-timing ASICs like the ITkPix~\cite{Alimonti:RD53:2025} or Timepix~\cite{Poikela:Timepix3:2014} chips.

\paragraph{Required infrastructures}
Muography applications can be operated parasitically.

\subsubsection{Testbed for a Muon Collider Demonstrator}
A multi-TeV muon collider has the unique potential to provide both precision measurements and the highest energy reach in one machine. One of the key challenges in its development is the delivery of a high brightness muon beam, which is essential to produce sufficient luminosity. Ionization cooling is currently the only feasible option for cooling a muon beam. In the muon collider current design, the injected muons will undergo ionization cooling, passing through 12 cooling stages, each stage consisting of about 100 identical cells. Each cell is composed of an absorber, an RF cavity, and a magnetic lens. Cells in different stages vary in complexity, starting with a simple magnetic arrangement and evolving to a much more complicated configuration at the end of the cooling channel. Although MICE proved the physics principles of ionization cooling,the challenges associated with the cooling technology and its integration remain the bottleneck. To understand and mitigate these risks, a demonstrator facility containing a sequence of ionization cooling cells, closely resembling a realistic cooling channel, is required. Such a facility will not only enable the design and testing of each component of a cooling cell but will also allow the integrated performance of the full system to be demonstrated, ensuring that no showstoppers exist for such technologies. 

Considering that the muon injector is expected to deliver 200-MeV muons at the front end, followed by a cooling section to reduce the emittance of the beam before acceleration, the Jefferson Lab secondary beam could serve as a testbed for the first cooling sections at selected energy intervals from 100\,MeV to a few\,GeV and for supporting the development of the ancillary equipment necessary for beam diagnostics.




\section{Neutrino Beams and Physics}
%
\subsection{Neutrino Beams at the Beamdump Facility}
%
The study of neutrino interactions has become a cornerstone of modern particle physics, with direct implications for our understanding of the SM, the properties of nuclear matter, and possible BSM physics. The next generation of long-baseline neutrino experiments, such as DUNE and Hyper-K, will rely on precise knowledge of neutrino--nucleus interactions over a broad energy range. Achieving this precision requires both high-statistics measurements and careful control of systematic uncertainties. Jefferson Lab provides unique opportunities for this program. Its continuous-wave high-intense electron beam, combined with the planned beamdump facility, can serve as an intense and clean secondary source of neutrinos. Primary $e^-$ interactions in the Hall~A dump generate secondary $\pi^+$ and $K^+$ that are stopped before decaying and produce well-characterized neutrino fluxes from \emph{pion decay at rest} ($\pi$--DAR) and \emph{kaon decay at rest} ($K$--DAR)~\cite{Battaglieri:2023gvd}. $\pi$--DAR produces a prompt ($\tau_\pi \!\simeq\! 26$\,ns) monoenergetic $\nu_\mu$ at $E_\nu = 29.8$\,MeV from $\pi^+ \!\to\! \mu^+ \nu_\mu$, and delayed ($\tau_\mu \!\simeq\! 2.2\,\mu$s) $\nu_e$ and $\bar{\nu}_\mu$ from $\mu^+ \!\to\! e^+ \nu_e \bar{\nu}_\mu$ with energies up to $\sim$ 50\,MeV. $K$--DAR yields a monoenergetic $\nu_\mu$ at $E_\nu = 236$\,MeV from $K^+ \!\to\! \mu^+ \nu_\mu$ (BR $\sim 65\%$). These well-understood spectra enable precision SM measurements and sensitive BSM searches, and their reach should be evaluated in the context of complementary proton beamdump based stopped-meson sources worldwide (e.g., at ORNL, LANL, FNAL, J-PARC).

\begin{figure}[htbp]
  \centering
  \includegraphics[width=0.48\textwidth]{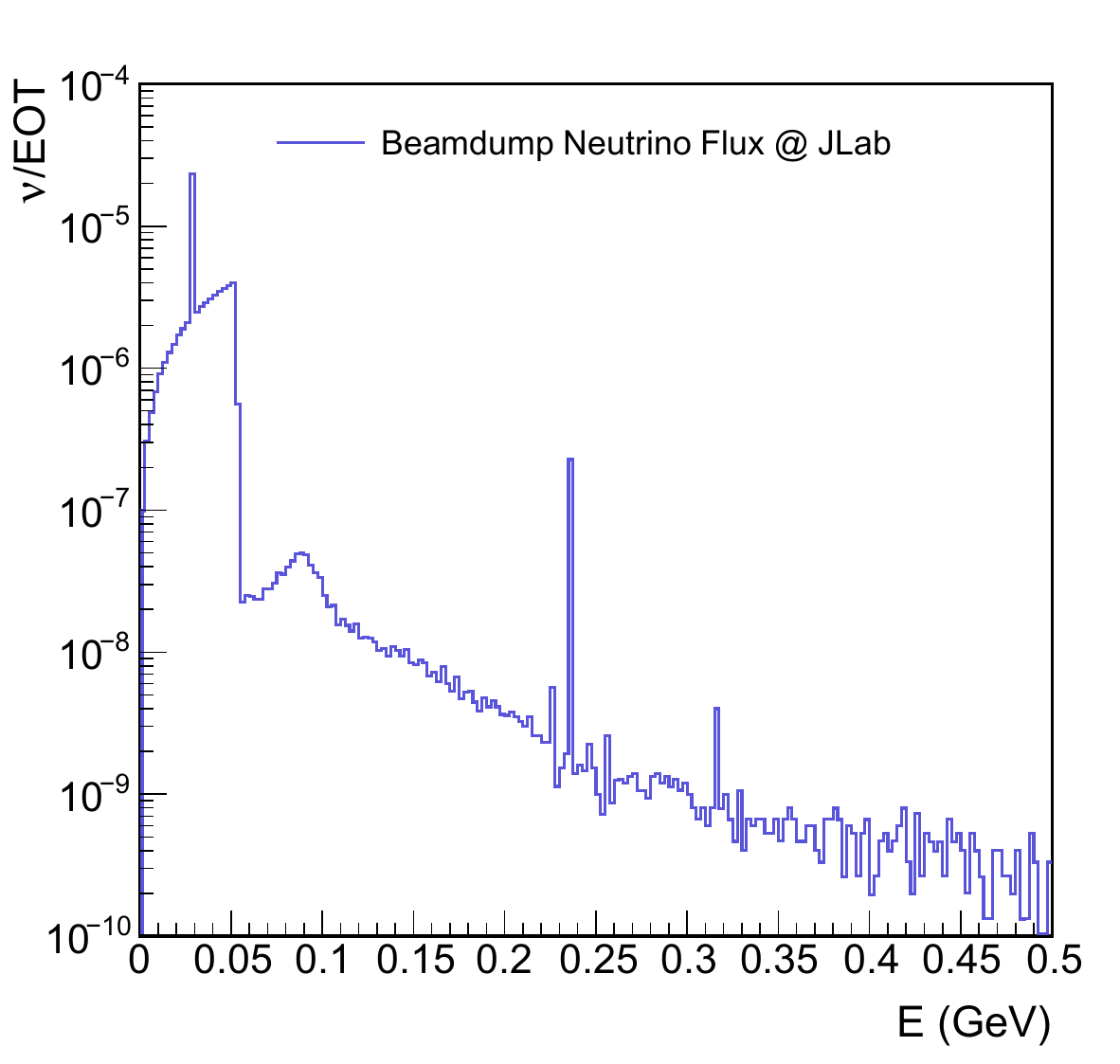}
  \includegraphics[width=0.48\textwidth]{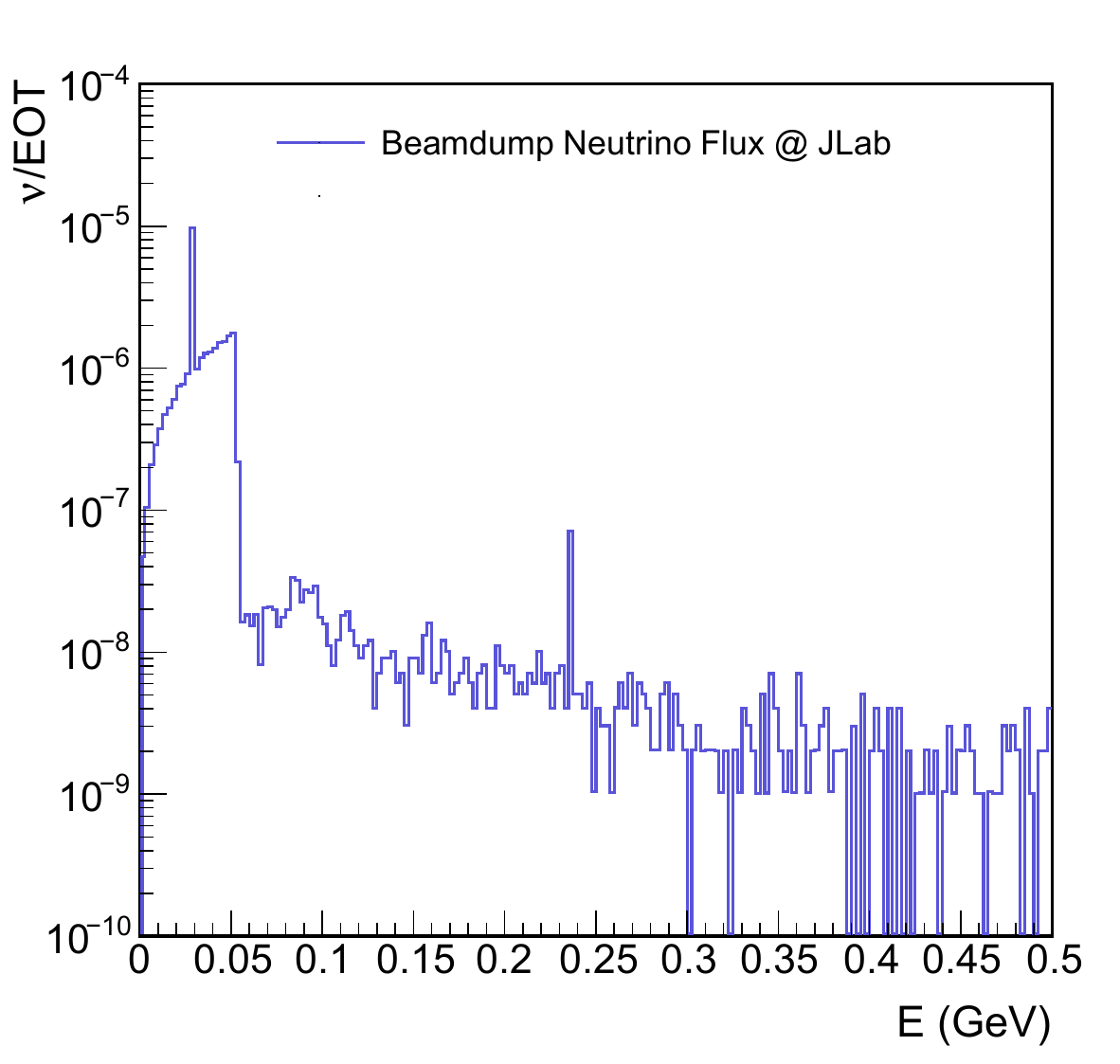}
  \caption{Energy distributions of neutrinos produced by the interaction of an 11-GeV electron beam with the Hall-A beamdump. {\bf Left:} Off-axis $\sim$ 10\,m above the dump. {\bf Right:} On-axis $\sim$ 5\,m downstream of the dump.}
  \label{fig:neutrino_11GeV}
\end{figure}

To characterize this source, realistic Monte Carlo simulations were carried out to evaluate the neutrino flux generated by the interaction of an 11-GeV, 50-$\mu$A electron beam with the Hall-A beamdump. The flux was sampled at two locations: an off-axis position approximately 10\,m above the dump and an on-axis position downstream of the dump. Figure~\ref{fig:neutrino_11GeV} shows the corresponding neutrino energy spectra.

The results indicate that the off-axis neutrino spectrum is consistent with that expected from a DAR source. In this configuration, the total neutrino flux in the energy range 0--100\,MeV is about $6.6 \times 10^{-5}\,\nu/$EOT, accounting for $\sim 99\%$ of the spectrum. For the on-axis configuration, although the DAR contribution remains dominant, a small but non-negligible high-energy component ($E_\nu > 100$\,MeV) is observed. The total on-axis neutrino flux in the range 0--500\,MeV is approximately $2.9 \times 10^{-5}\,\nu$/EOT, with the DAR component representing $\sim 96\%$ of the overall yield. The resulting flux of electron and muon neutrinos, peaking below 55\,MeV and extending up to $\sim$ 200\,MeV, provides an optimal environment to explore coherent elastic neutrino--nucleus scattering (CEvNS), search for LDM, and test neutrino event generators in a new regime.

\subsection{Neutrino Detection at the Beamdump Facility}

\subsubsection{Using a Liquid Argon Detector}
The expected neutrino interaction rates range from several tens to few hundreds of electron neutrino charge current and from several hundred to a few thousand coherent neutrino scattering events per year per ton of target material positioned in proximity of the beamdump. The liquid Ar time projection chamber is a well-established technology conceived and widely used for neutrino physics and rare event searches in the energy range from few MeV to several GeV. The double phase liquid Ar time projection chamber technology was specifically developed to identify low energy nuclear recoils (from few to several ten keV Ar recoil energy) produced by interaction of DM in the form of WIMPs; it is therefore perfectly suited to detect coherent neutrino scattering events and its dynamic energy range could be easily extended to a few tens of MeV.
  
\begin{figure}[htbp]
  \centering
  \includegraphics[width=0.48\textwidth]{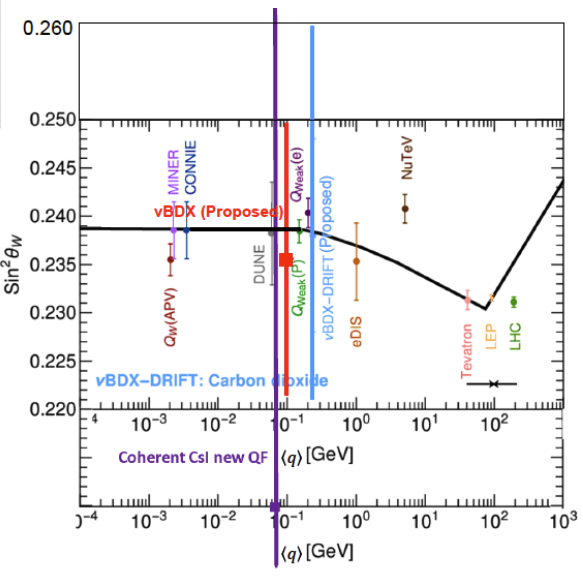}
  \caption{Projected uncertainty for a CEvNS measurement of the Weinberg angle ($\sin^2\theta_W$) obtained from one year of exposure of a 1\,m$^3$ liquid Ar detector operated at a 10\,keV energy threshold. The total error includes systematic uncertainties, dominated by the beam-related neutron background, assumed to be mitigated by $\sim 1$\,m of lead shielding placed beneath the detector.}
  \label{fig:CEvNS}
\end{figure}

First estimates show that, in a year of exposure, thousands of CE$\nu$NS events could be recorded in a $\sim 1$~ton liquid Ar detector operating at a $\sim 10$~keV energy detection threshold, located above the beamdump. This would allow, for instance, the extraction of $\theta_W$ with a precision at the level of $\sim 10\%$ (including systematic uncertainties dominated by the neutron background), which represents a far better accuracy than the current extraction performed by the COHERENT experiment, based on only a few hundred events. Figure~\ref{fig:CEvNS} illustrates the potential of a CEvNS measurement at Jefferson Lab. Despite the fact that this would not be competitive with the projected precision of the MOLLER experiment, it would  provide an important cross-check of the measurement in a completely independent way and in an unexplored kinematic region. Further studies are necessary to optimize the shielding and evaluate the sensitivity of such a measurement to other physics observables (e.g., BSM physics).
   
\paragraph{Targets and instrumentation required} 
Considering the significant R\&D effort carried out by leading multi-ton experiments that use this technology, such as DarkSide and DUNE, liquid argon prototype detectors with active target masses ranging from tens to hundreds of kilograms are already available for test experiments. The detector could be installed in a suitable location off-axis near the beamdump to maximize the exposure to the secondary low energy neutrino beam while minimizing the high energy secondary particles produced in the dump. A few meters of concrete would provide the minimal required shielding against cosmic rays and beam-related background. Such an exploratory measurement would provide significant insight before planning for a full-scale experiment in a optimized facility.

\paragraph{Required infrastructures}
After a pilot experiment performed above ground, which would require only minimal dedicated infrastructure (e.g., a shelter with concrete shielding for the prototype, power, and air conditioning), a new underground dedicated hall adjacent to the vault would be needed. The footprint of a liquid argon-based detector is estimated to be 5--10\,m$^2$, including space for readout electronics. Cryogenic systems, such as recirculation and purification, are required for operation. The typical power consumption for few-ton-scale detectors is below 50\,kW. The thickness of the neutron shielding must be evaluated based on the neutron energy spectrum and the chosen shielding material.


\subsubsection{Using Opaque Liquid Scintillators}

A promising technology for neutrino detection involves the use of opaque liquid scintillators read out by optical fibres, a novel class of scintillating detectors where the scintillating liquid is made optically opaque by adding materials like waxes~\cite{Buck_2019}. Key features are
(i) light transport by scattering 
and (ii) segmentation without physical barriers. 
Further advantages are reduced optical cross-talk, good spatial reconstruction, and potential cost savings in the case of large volumes. Applications include particle detectors where the topology of the interaction and timing are important. 

A new technological advancement for reducing light losses in wavelength-shifting light guides are fibers 
where the wavelength-shifter material is located near the outer surface. Photons that are absorbed and emitted on the outer surface have a higher chance of being captured by total internal reflection compared to those closer to the center of the fiber~\cite{OWL2}. Recently, it was demonstrated that detectors can achieve few-mm spatial resolution~\cite{liquido1}. 

\begin{figure}
  \centering
  \includegraphics[width=0.9\textwidth]{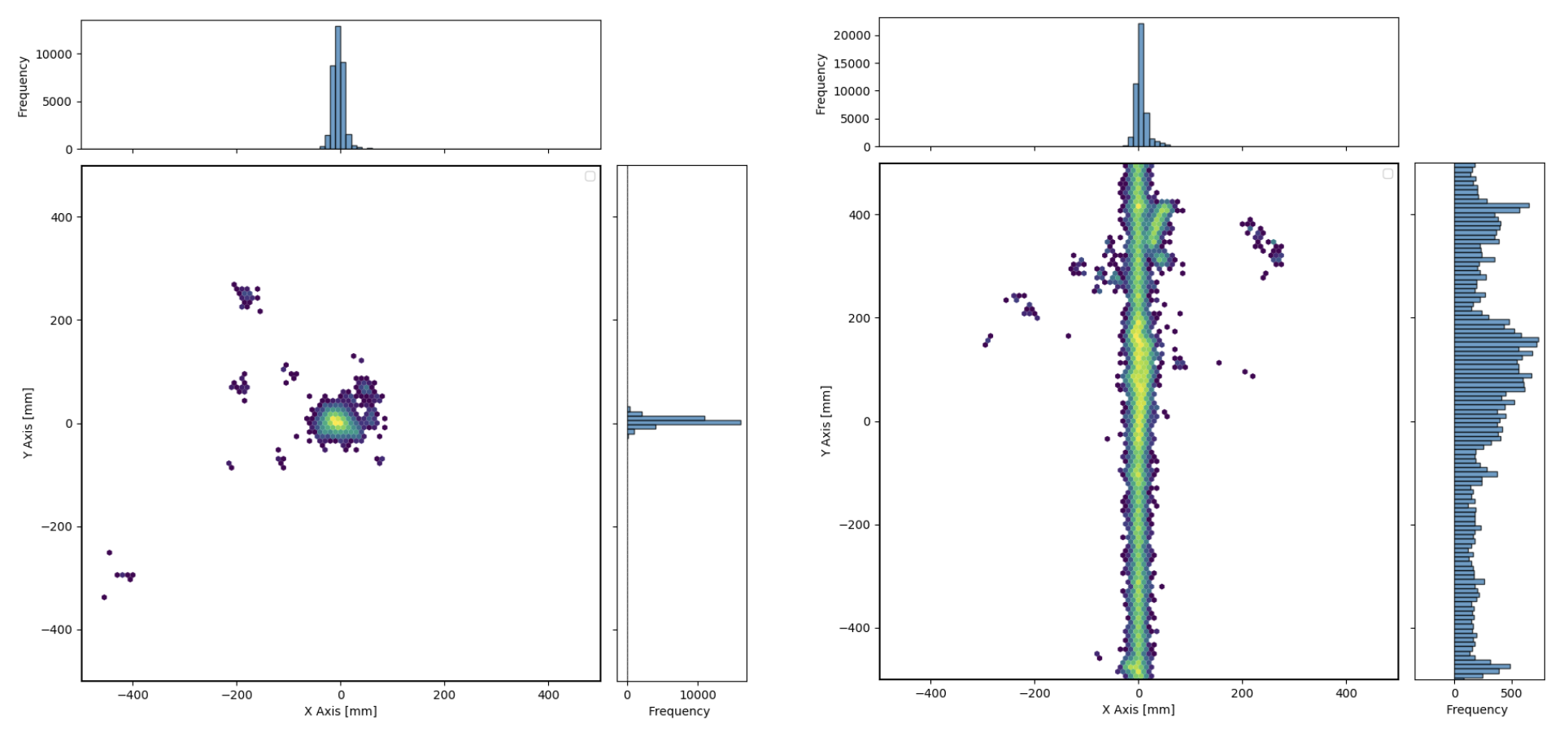}
  \caption{Simulation of the light signal for 2-GeV muons entering a 1\,m$^3$ opaque liquid scintillator detector with fibers spaced at 1\,cm. {\bf Left:} Muon entering the detector from above parallel to the fiber direction. {\bf Right:} Muon entering from the front face orthogonal to the fiber direction. Shown are selected events in which secondary interactions are visible.}
  \label{fig:OLD}
\end{figure}

An additional detector based on opaque scintillators would have the potential to improve the sensitivity of DM searches to lower masses by lowering the energy threshold and complementing the BDX program. The tracking capabilities of such a detector can also help reject low-energy cosmic-ray–induced backgrounds through reconstruction of the event spatial topology. The simulation of 2~GeV muons detected by the proposed detector design is shown in Fig.~\ref{fig:OLD}. 

The neutrino energy range at Jefferson Lab is well suited for liquid scintillators in general. 
However, considering a 1\,m$^3$ detector volume filled with an opaque liquid scintillator with density $\rho\approx 1$\,g$/$cm$^3$ and a neutrino interaction cross section in the MeV range of $10^{-43}$\,cm$^2$, fewer than one interaction per year on hydrogen and approximately two interactions per year on carbon are expected. Therefore, for neutrino applications in the MeV range, a larger detector volume is necessary. Such a detector could be placed off-axis with respect to the beamdump. In addition to enabling detection of CEvNs on hydrogen/carbon, such a detector could be used for studying specific neutrino--nucleus reaction channels for improving $\nu$--nucleus interaction models in the $\mathcal{O}$(100\,MeV) range for applications to oscillation experiments with beams from kaon decays at rest~\cite{Nikolakopoulos:2020alk}.



\paragraph{Targets and instrumentation required}
  The detector could consist of a $\sim$ 1\,m$^3$ tank containing the opaque scintillator, enclosed within an outer tank filled with  transparent scintillator, resulting in a total footprint of approximately 1.7\,m$^2$. Additional space of about $\sim$ 1\,m$^2$ is needed for scintillator preparation and mixing. Furthermore, plastic scintillators may be required as additional cosmic-ray veto detectors.
\paragraph{Required infrastructure}
  Considering a 1\,cm spacing between the fibers and dual readout, we estimate a total of 20,000 channels. The light collected by the fibers is detected by silicon photomultipliers (SiPMs). The main contributor to power consumption is the waveform digitizers (e.g., CAEN 1740 modules consuming approximately 31\,W per module with 64 channels each). Including SiPM biasing, pre-amplifiers, DAQ computers, cooling, and racks, the total estimated power consumption is about 26\,kW. Power consumption and rack footprint could be reduced in future designs by using denser modules or dedicated ASIC readout.
  The infrastructure could be shared with a LAr detector of similar size and the two could run in parallel.



\subsection {Prospects for Neutrino Physics at the Beamdump Facility}

\begin{figure}[htbp]
  \centering
  \includegraphics[width=0.96\textwidth]{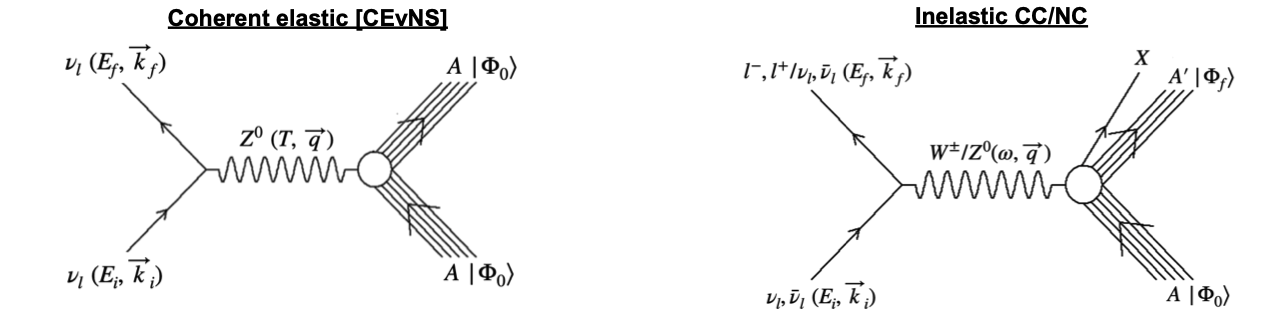}
  \caption{Neutrino--nucleus interactions. {\bf Left:} Coherent $\nu$-nucleus elastic scattering. {\bf Right:} Inelastic scattering.}
  \label{fig:nu-iteractions}
\end{figure}

The neutrino flux from $\pi$-- and $K$--DAR spans tens to hundreds of MeV, enabling access to a broad range of interaction channels. At $\pi$--DAR energies ($E_\nu \lesssim 50$~MeV), coherent elastic neutrino--nucleus scattering (CEvNS) dominates (Fig.~\ref{fig:nu-iteractions}-left). This process, in which the neutrino scatters coherently off the entire nucleus, benefits from an enhancement proportional to the square of the neutron number and leads to a characteristic nuclear recoil in the $\mathcal{O}(10-100)$~keV range. Theoretical uncertainties on CEvNS cross section are estimated of the order of a few percent, supporting precision SM tests and sensitivity to BSM signals~\cite{Pandey:2023arh, Tomalak:2020zfh}. 

Inelastic channels (Fig.~\ref{fig:nu-iteractions}-right), both charged-current and neutral-current, are also accessible. For $\pi$--DAR energies, these probe nuclear excitations with $\mathcal{O}(10$\,MeV) $\gamma$, $p$, and $n$ emissions. These processes are complementary to neutrinos from the core-collapse supernova, whose energies overlap with the pion decay at rest neutrinos. Such cross sections remain poorly known, with limited measurements and no existing data for the argon nucleus relevant for DUNE's supernova program. For $K$--DAR at 236~MeV, interactions occur in the challenging transition between collective nuclear excitations and quasielastic scattering off bound nucleons~\cite{Nikolakopoulos:2020alk}

This physics program enables a range of SM measurements, $\pi$--DAR CEvNS enables precision extractions of nuclear weak form factors and neutron radii, complementing parity-violating electron scattering experiments at Jefferson Lab, and measurements of $\sin^2\theta_W$ at low $Q^2$, complementing the MOLLER experiment. $\pi$--DAR inelastic scattering measurements would shed light on low-energy nuclear structure that would complement supernova neutrino studies. $K$--DAR scattering would provide valuable constraints on the axial component of the neutrino–nucleus interaction at low energies, using monoenergetic beams to enable missing-energy determinations~\cite{JSNS2:2024bzf}. 

From the BSM perspective, CEvNS offers a powerful probe of new physics: deviations from the SM prediction, whether in the total event rate or in the shape of the nuclear recoil spectrum, would signal additional contributions to the cross section. These can arise from neutrino electromagnetic properties (neutrino magnetic moment and neutrino charge radius), non-standard interactions of neutrinos, or even be sensitive to sterile neutrino oscillations~\cite{Pandey:2023arh, Tomalak:2020zfh}. 

The modeling of neutrino--nucleus interactions is a leading source of uncertainty in oscillation measurements. Upcoming long-baseline neutrino experiments such as DUNE and Hyper-K have dedicated programs to measure relevant cross sections~\cite{DUNENearDetector,HyperKDesignReport}, though these efforts are complicated by the broad energy distribution of neutrino beams. As discussed in the previous section, lepton--nucleus scattering measurements provide a complementary approach, probing key aspects of nuclear dynamics with charged beams and high statistics. 


\section{Synergies and Funding Opportunities}
%
Experiments with secondary beams at Jefferson Lab, building upon the approved LDM searches, would significantly extend the current physics program of the laboratory. This would place Jefferson Lab in line with other world-leading facilities such as Fermilab, TRIUMF, CERN, PSI, J-PARC, and RIKEN, where secondary beams already enrich their primary missions. Pilot experiments could be carried out with minimal additional infrastructure, attracting new users from diverse physics communities (basic science, high-energy physics, applied physics, condensed matter) and enhancing the laboratory’s role at the national level. Upon verification of the potential of an opportunistic secondary beam program at Jefferson Lab, dedicated and optimized infrastructures could be designed and deployed to take full advantage of a world-competitive facility.  

A diversified scientific program would be of significant interest both to institutions already engaged in Jefferson Lab’s scientific program and to new partners willing to contribute. For instance, Collaborative NSF grants with Primarily Undergraduate Institutions (PUIs) for the beamdump facility represent an opportunity to broaden participation in nuclear physics. PUIs play a critical role in advancing scientific education and research by fostering hands-on learning and mentorship opportunities for students. They can also contribute to the beamdump facility through participation in NSF Major Research Instrumentation (MRI) grants. By focusing on detector component development and testing, or on software development, PUIs can contribute meaningfully to large-scale experiments while providing transformative research opportunities for undergraduates. These activities not only advance the experiment but also immerse students in authentic research, aligning with the NSF’s goal of broadening participation in STEM~\citep{NSF2022}. The success of new institutional involvement will hinge on collaboration between new and established research institutions, thereby strengthening Jefferson Lab’s partnerships. An NSF MRI grant can further enable a consortium approach, where multiple institutions pool resources to support a broad scientific program. Shared instrumentation and expertise would allow each partner to focus on complementary aspects of detector development. Funding opportunities will not be limited to NSF or DOE; other agencies already engaged at the laboratory, as well as new international collaborators, could also support a secondary beam program. 

\section{Summary}
%
A beamdump facility at Jefferson Lab offers unique and pioneering opportunities once constructed. Besides the light dark matter searches carried out by the Beam Dump eXperiment (BDX), intense muon and neutrino beams over a wide energy range could be exploited to broaden the current scientific program of the laboratory. 

A secondary muon beam with energies from 100\,MeV to a few\,GeV would be competitive with the highest intensity beams available in the US and overseas, providing unique features such as GeV-energy fully polarized muons. These could strengthen the current nuclear physics program (e.g., precise studies of QED effects in electron scattering, nucleon form factor determination), offer new opportunities in BSM physics (e.g., muon--electron scattering), support applied physics (e.g., muon tomography and elemental analysis), and serve as a testbed for R\&D of future muon-collider components.  

Jefferson Lab’s secondary neutrino beam program represents a unique and timely opportunity. By combining intense neutrino fluxes with a characteristic DAR energy spectrum and advanced low-threshold detectors (liquid Ar or novel scintillating materials), it can deliver precision measurements of Coherent Elastic Neutrino-Nucleus Scattering (CEvNS), a recently confirmed process that provides information on fundamental observables such as the Weinberg angle, opens new opportunities for BSM searches, and delivers critical input for long-baseline oscillation experiments (e.g., DUNE). Leveraging the existing infrastructure and expertise, Jefferson Lab is positioned to become a leading hub for intensity-frontier neutrino physics.  

Although not the subject of the workshop, opportunities with high-intensity secondary neutron beams have also been explored. Preliminary results indicate that, for certain energy ranges, Jefferson Lab's neutron beam could be competitive. A detailed discussion of secondary neutron beams at Jefferson Lab will be addressed in a future edition of this workshop.

In conclusion, with support from experts at leading worldwide facilities with muon and neutrino physics programs, the {\em International Workshop on Secondary Beams at Jefferson Lab: BDX \& Beyond} has shown that a secondary beam program at Jefferson Lab would enhance the current scientific program, offering top-quality muon and neutrino beams. This program could attract a diversified community from high-energy, basic science, applied, and condensed matter physics, complementing the laboratory's leading role in nuclear physics. Leveraging the underground vault planned for light dark matter searches, opportunistic pilot experiments and demonstrators can be deployed efficiently. Upon verification, optimized infrastructures would further reinforce Jefferson Lab’s strategic role as a unique, multi-purpose, world-leading accelerator facility.

\section*{Acknowledgment}
This material is based upon work supported by the U.S.\ Department of Energy, Office of Science, Office of Nuclear Physics under contract DE-AC05-06OR23177.



\end{document}